\documentclass[floats,prd,aps,showpacs,amssymb,nofootinbib]{revtex4}
\usepackage{amssymb}
\usepackage{graphics}
\usepackage{amsmath}
\usepackage{amsfonts}
\usepackage{bm}% bold math
\usepackage[]{latexsym}

\newcommand{\mlab}[1]{\label{#1}}
\newcommand{\EX}{{$\eta$-$\xi$ }}

\newcommand{\etad}{\eta_{_\delta}}
\newcommand{\xid}{\xi_{_\delta}}
\newcommand{\I}{{_I}}
\newcommand{\II}{{_{I\!I}}}
\newcommand{\III}{{_{I\!I\!I}}}
\newcommand{\IV}{{_{I\!V}}}
\newcommand{\IId}{{_{I\!I_{\!\delta}}}}

\newcommand{\IQ}{{_I,_{I\!I}}}
\newcommand{\R}{{\Bbb R}}
\newcommand{\Vk}{{\bf k}}
\newcommand{\Vp}{{\bf p}}
\newcommand{\Vx}{{\bf x}}

\usepackage{xcolor}
\usepackage{graphicx}
\usepackage{amssymb}

\begin{document}

\title{Unified formalism for Thermal Quantum Field Theories: a geometric viewpoint}

\author{M.~Blasone$^{1,2,}\footnote{blasone@sa.infn.it}$, P.~Jizba$^{3,4,}\footnote{p.jizba@fjfi.cvut.cz}$ and G.~ G.~Luciano$^{1,2,3,}\footnote{Corresponding author: gluciano@sa.infn.it}$}

\affiliation{\mbox{$^1$Dipartimento di Fisica "E.R. Caianiello", Universit\'a di Salerno, Via Giovanni Paolo II, 132 I-84084 Fisciano (SA), Italy.} \\ 
  \mbox{$^2$INFN, Gruppo collegato di Salerno, Italy.}\\ 
  \mbox{$^3$FNSPE, Czech Technical University in Prague, B\v{r}ehov\'{a} 7, 115 19 Praha 1, Czech Republic}
  \mbox{$^4$ITP, Freie Universit\"{a}t Berlin, Arnimallee 14, D-14195 Berlin, Germany}
  }

%\author[Sa,Infn]{M.~Blasone}
%\ead{blasone@sa.infn.it}

%\author[cvut,ger]{P.~Jizba}
%\ead{p.jizba@fjfi.cvut.cz}

%\author[Sa,Infn,cvut]{G.G.~Luciano\corref{cor1}}
%\ead{gluciano@sa.infn.it}
%\cortext[cor1]{Corresponding author}

%\address[Sa]{Dipartimento di Fisica, Universit\`a di Salerno, Via Giovanni Paolo II, 132, 84084 Fisciano, Italy}
%\address[Infn] {INFN Sezione di Napoli, Gruppo collegato di Salerno, Italy}
%\address[cvut]{FNSPE, Czech Technical University in Prague, B\v{r}ehov\'{a} 7, 115 19 Praha 1, Czech Republic}
%\address[ger]{ITP, Freie Universit\"{a}t Berlin, Arnimallee 14, D-14195 Berlin, Germany}

\begin{abstract}
In this paper we study a unified formalism
for Thermal Quantum Field Theories, i.e., 
for the Matsubara approach, Thermo Field Dynamics 
and the Path Ordered Method. To do so, we employ a 
mechanism akin to the Hawking effect which 
explores a relationship between the concept of 
temperature and spacetimes endowed with event-horizons.
In particular, we consider an eight dimensional static 
spacetime, the so-called  \EX spacetime, which we show to 
form an appropriate geometric background for generic Thermal 
Quantum Field Theories. Within this framework, the different 
formalisms of Thermal Field Theory are unified in a very natural 
way via various analytical continuations and the set of time-paths 
used in the Path Ordered Method is interpreted in geometric 
terms. We also explain reported inconsistencies inherent in 
the Thermo Field Dynamics through the appearance of horizons 
(and ensuing loss of information) in the \EX spacetime.
\end{abstract}

%\begin{keyword}
%Thermal Field Theories\sep$\eta$-$\xi$ spacetime
%\end{keyword}

%\end{frontmatter}
 \vskip -1.0 truecm
\maketitle

\section{Introduction}
%%%%%%%%%%%%%%%%%%%%%%%%%%%
\setcounter{equation}{0}
Under the generic name \emph{Thermal Quantum Field Theory} (TQFT)~\cite{DAS,LEB,LW87,KAP,BJV,UM0} one collects all formalisms of Quantum Field
Theory (QFT) at finite temperature and density, i.e., the Matsubara or Imaginary Time (IT) approach~\cite{DAS,LEB,LW87},
Thermo Field Dynamics (TFD)~\cite{DAS,UM0,UM01,UM1} and the Path Ordered Method (POM)~\cite{BJV,Mills}. The latter includes, for instance, the familiar Closed Time Path (CTP) formalism of Keldysh and Schwinger~\cite{DAS,LEB,LW87} as a special case.
The existence of these distinct approaches results from conceptually different efforts to introduce a temperature within the framework of QFT. For instance, in the IT formalism one exploits the analogy between the (inverse) temperature and imaginary time in calculating the partition function. Within this approach, the two-point Green function is given by the Matsubara propagator and the sum over Matsubara frequencies (alongside with various summation techniques) must be invoked when dealing with multi-loop thermal diagrams. Because time is traded for (inverse) temperature, the IT formalism cannot directly address field dynamics within a heat bath and it is thus suitable basically only for QFT at thermal equilibrium.
In principle, real-time quantities can be obtained by analytic continuation to the real axis, but in practice this procedure is plagued with ambiguities and further delimitations are typically needed~\cite{LEB,LW87,Dolan}. Aforementioned ambiguities typically appear in higher-point Green''s functions, e.g., in three-point thermal Green's functions~\cite{Baier,Evans,Frenkel}. Ambiguities were also reported in the $\beta$-function calculations at the one-loop level~\cite{Braten,LandsmanIII}.
%The basic disadvantage of the Matsubara
%formalism lies in the unphysical representation of time and energy.
%*** plasmons.

In contrast, both the POM formalism and TFD accommodate time and temperature on equal footing and no extra analytical continuation is required. On one hand, this provides a powerful theoretical platform allowing to address such issues as a temporal dynamics of quantum fields in a thermal heat bath~\cite{PVL,GUI92ferm,JT}, dynamics of phase-transition processes~\cite{BJV} or linear responses~\cite{LEB}. On the other hand, the price to be paid for working with the real time is the doubling of the field degrees of freedom which is reflected in a $2 \times 2$ matrix structure of thermal propagators and self-energies. Consequently, in higher-loop orders a much larger number of diagrams has to be taken into account as compared to the vacuum (i.e., zero temperature) theory. Surprisingly, the POM and TFD approaches lead to the same matrix form for thermal propagator in equilibrium~\cite{H95}. This is a quite intriguing fact since  POM and TFD  have very different conceptual underpinnings.
In the POM formalism one introduces the temperature by adding a pure imaginary number to the real time and chooses a special time path in the complex-time plane which involves the use of both time- and anti-time-ordered Green functions. In comparison with this, the TFD
is a method for describing mixed states as pure states in an enlarged Hilbert space (akin to a purification procedure used in quantum optics~\cite{Cirac}).
It is characterized by a doubling of the field algebra and and its mathematical underpinning is provided by algebraic quantum field theory ($C^*$-algebras,
the Haag--Hugenholtz--Winnink (HHW) formalism and Tomita--Takesaki modular theory). In the TFD  the temperature is contained explicitly in the resulting ``vacuum''
pure state, which is referred to as the ``thermal vacuum'' state. Ensuing field propagators are then expressed as expectation values of time-ordered products of quantum
fields with respect to such a thermal state. From the above considerations, it appears that a doubling of the degrees of freedom is necessary in order to be able to calculate real-time
Green functions. This doubling is, however, absent in the Matsubara formalism, and thus it could a mere mathematical artifact. Yet, such a doubling of the field degrees of freedom
also appears in the axiomatic formulation of quantum statistical mechanics~\cite{Haag,HaagII,Sewel}.
This indicates that a two-component extension is essential for a consistent Minkowski-space field theory at finite temperature and density. This viewpoint will be also confirmed
by our subsequent analysis.

An interesting question is whether the aforesaid TQFT formalisms have some
roots in common or, in other words, if their features can be understood in a deeper way
so that they appear to be unified. A clue to an answer may be found in the well-known
discovery of Hawking~\cite{HAW} that temperature may arise in a quantum theory as a
result of a non-trivial background endowed with event-horizon(s).
In Hawking's case the background in question is any
asymptotically flat black-hole spacetime, such as  Schwarzschild~\cite{HAW,Page},
Reissner--Nordstr\"{o}m~\cite{PageII,Sorkin} or Kerr~\cite{PageI} spacetime.
Rindler spacetime, i.e., the spacetime of an accelerated observer, is also known to exhibit thermal features~\cite{Rin,BIDA,Blasone:2017nbf,Blasone:2018byx}.
This fact is known as the Unruh (or Davies--Unruh) effect~\cite{Unruh:1976}. The same logical scheme can be extended also to cosmological
horizons, like the event horizon in de Sitter spacetime, where the ensuing (de Sitter) temperature is
$T_{\rm dS}=H/2\pi$ ($H^{-1}$ is the radius of the horizon -- de Sitter radius)~\cite{Guo,Bousso}.

In all these cases the (zero-temperature) vacuum state of an inertial observer is perceived as a thermal state
by a certain kind of ``non-inertial'' observers, e.g.,
a black-hole spacetime (Hartle--Hawking vacuum) represents a thermal state for a static (i.e., non-inertial) observer in Schwarzschild
spacetime~\cite{HAW,Israel:1976}, and similarly, Minkowski vacuum agrees with a thermal state for an accelerated (i.e.,
non-inertial) observer~\cite{Unruh:1976}. It has been known for some time that the aforesaid
concept of an observer-dependent vacuum, or more precisely, an observer-dependent notion of particle being
emitted from the horizon (alongside with the ensuing concepts of a heat bath and temperature) offers an interesting
route towards unification of some of TQFT formalisms~\cite{GUI90}.
The merger can be achieved when one constructs a new spacetime in which the (zero-temperature) vacuum corresponds to a usual
thermal state for Minkowski inertial observer, i.e, where the Minkowski observer
is an appropriately chosen non-inertial observer from the point of view of the new spacetime.
In addition, such a spacetime should have more than 4 dimensions to allow for analytical continuation between
Minkowski and Euclidean spacetimes~\cite{GH}.
Along these lines, a flat background with a non-trivial horizon structure providing desired thermal features -- the so-called \EX spacetime -- has been constructed~\cite{GUI90,ZG95,ZG98}.

In its essence, the \EX spacetime is a flat complex manifold with complexified $S^1\times R^3$ topology.
Its Lorentzian section consists of four copies of Minkowski spacetime glued together
along their past or future null hyperplanes. Since in Kruskal-like coordinates the metric is singular
on these hyperplanes, we shall call them formally event-horizons~\cite{note}. Their existence leads to the doubling of the
degrees of freedom of the fields. The vacuum propagator on this section is found to be equal to
the real-time thermal matrix propagator. On the other hand, in the Euclidean section of
\EX spacetime, the time coordinate is periodic and the field
propagator can be identified with the conventional Matsubara propagator.
%
%*** The vacuum of quantum fields in the \EX spacetime is just the thermal state
%for an inertial observer in Minkowski or Euclidean spacetime.

%Our aim here is to show that the \EX spacetime is a generic example of a background exhibiting thermal
%features in the sense that fields and states in this spacetime look
%everywhere as if they were immersed in a thermal bath contained in a
%Minkowski background~\cite{GUI90}.

\begin{figure}[t]
\begin{center}
\includegraphics[scale=0.45]{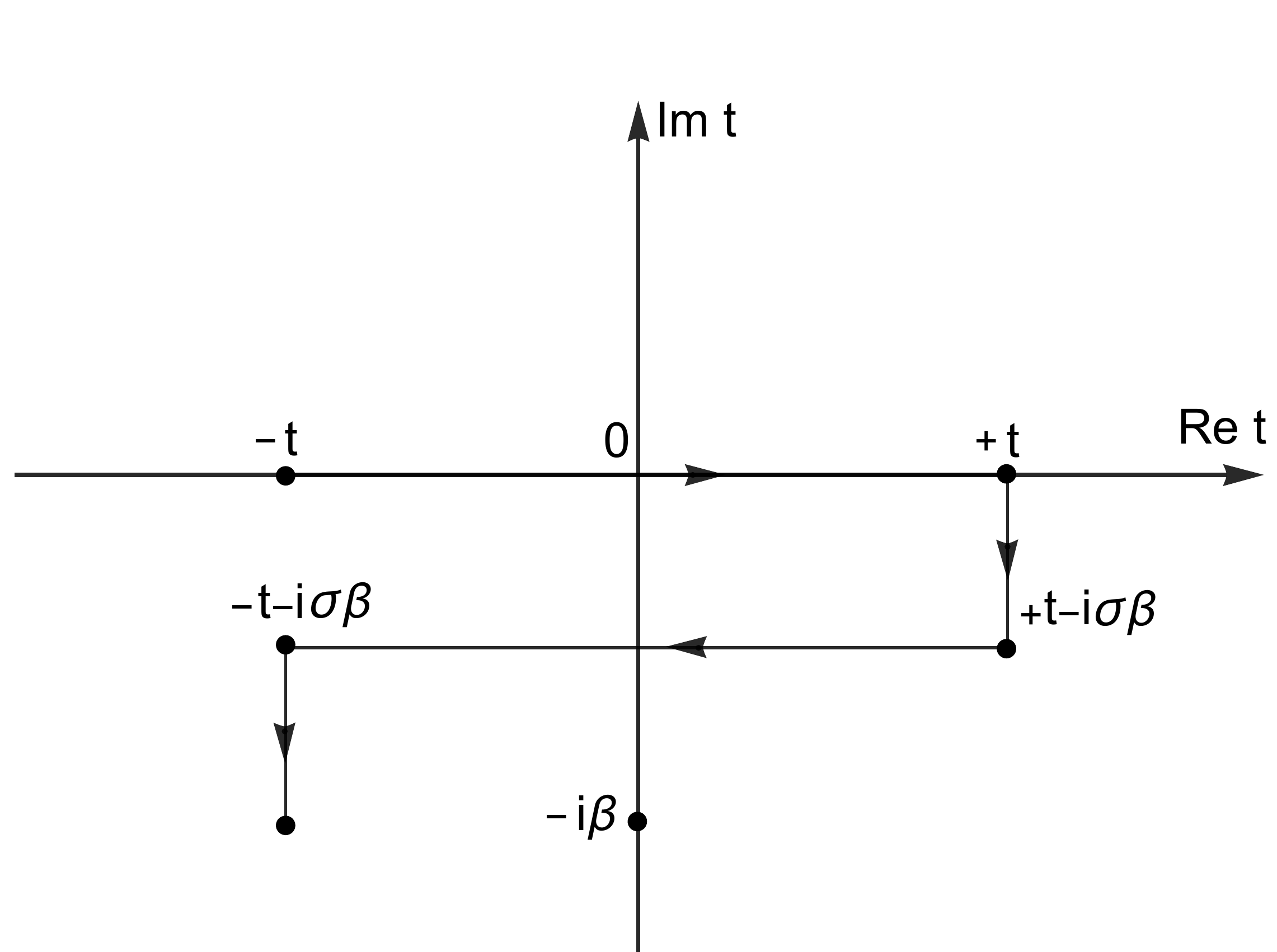}
\end{center}
\caption{The Niemi--Semenoff time path used in POM. The parameter $\sigma$ ranges from
the value $\sigma=0$ (Closed Time Path) to $\sigma=1$.}
\label{fig1}
\end{figure}

Our aim here is to show that the \EX spacetime is structurally richer than previously thought and,
in doing so, we point out the existence and relevance of other complex sections of the \EX
spacetime aside from the already known Lorentzian and Euclidean ones. In this context, it is worth recalling that TQFT formalisms
have a number of physically equivalent (though technically distinct) parameterizations~\cite{LEB,LW87,KAP,BJV}.
For instance, the real-time TQFTs are characterized by a freedom in the parameterization of
the thermal matrix propagator. In the POM formalism, this freedom in parameterization is related to the
choice of a specific path in the complex-time plane going from $t=0$ to $t=-i\beta$, which is not
unique~\cite{NS}. This, so called Niemi--Semenoff time path, is depicted in Fig.~\ref{fig1}.
In TFD, different parameterizations of the Bogoliubov thermal matrix are permitted~\cite{H95}.
We stress that although the choice of the parameterization (both in POM and TFD) is irrelevant in {\em thermal equilibrium}
since it does not affect physical quantities~\cite{BJV,H95},
it plays a crucial r\^{o}le in non-equilibrium situations, where
the choice of a closed-time path in POM or the {\em left} and {\em right} statistical states in TFD relate to a particular
form of a transport equation~\cite{H95}.
As yet, such distinct parameterizations of TQFTs have not been considered in the context of the \EX spacetime.
Here we will see that all the aforementioned parameterizations can be realized within
the \EX spacetime framework and can be interpreted in purely geometrical terms.
The geometric picture for TQFTs is consequently enlarged.  We will also show how the
unification of the different formalisms of TQFT arises naturally
within this framework. The generalization introduced here could be useful
in order to extend the geometric picture of \EX spacetime to systems
out of thermal equilibrium, for which the typical choice of equilibrium parameterization is not convenient~\cite{H95}.

In passing, we mention that the idea of translating some features of a physical system
in terms of a geometric background  may be of interest also in other contexts than those considered here.
For instance, it has been highlighted that in the standard QFT on Minkowski spacetime, flavor mixing
relations hide a Bogoliubov transformation
that is responsible for the unitary inequivalence of flavor and mass
representations and their related vacuum structures~\cite{BV95,bosonmix}.
The same scenario has been recently investigated also for an accelerated
observer~\cite{Blasone:2017nbf,Blasone:2018byx}. In this context it has been
pointed out that the Bogoliubov transformation associated to flavor mixing and the one
arising from the causal structure of the Rindler spacetime combine symmetrically in the calculation of the modified Unruh spectrum, suggesting a possible geometric interpretation for the origin of flavor mixing. Similarly, geometric effects related to the existence of a minimal observable length have been shown to underpin the loss of thermality of Unruh radiation in the framework of Generalized Uncertainty Principle (GUP)~\cite{Nirouei:2011zz}.
The outlined geometrical background picture could also help to shed new light on such issues as 
the cosmological holography~\cite{Lloyd} or $\alpha$-vacua in AdS/CFT correspondence~\cite{Chamblin}.

The remainder of this paper is organized as follows. Section~2 provides theoretical
essentials of the \EX spacetime. In doing so, we focus our attention on Euclidean and
Lorentzian sections alongside with their respective complex extensions. As a tool for our analysis, in Section~3 we show how to perform field quantization in \EX
spacetime. To keep our discussion as transparent as possible, we consider only a scalar quantum field.  Section~4 is devoted to examining the relationships
between \EX spacetime and TQFTs, and to the unification of various TQFTs
in the framework of \EX spacetime. Furthermore, we discuss some of the salient features of the extended Lorentzian
section. Various remarks and generalizations are addressed in the concluding Section~5.

\section{\EX spacetime -- essentials and beyond }\label{sec.2}
%%%%%%%%%%%%%%%%%%%%%%%%%%%%%%%%%%%%%%%%%%%%%%%%%%%%%%%%%%%%%%%%

The \EX spacetime was originally introduced in Refs.~\cite{GUI90}. It represents a four-dimensional complex (i.e., eight-dimensional real)
manifold defined by the line element
\begin{equation}
 \mlab{guimetric}
ds^2 \, = \, \frac{-d\eta^2 \, + \,  d\xi^2}{\alpha^2\left(\xi^2-
\eta^2\right)} \, + \, dy^2 \, + \, dz^2\, ,
\end{equation}
where $\alpha =  2\pi/\beta$ is a real constant and
$(\eta,\xi,y,z)\in{\mathbb {C}^4}$.  In the following, we will use the symbol
$\xi^\mu = (\eta,\xi,y,z)$ to denote the entire set of \EX
coordinates, but for simplicity we will often drop the index
$\mu$ when no confusion with the space-like coordinate occurs.

%%%%%%%%%%%%%%%%%%%%%%%%%%%%%%%%
\subsection{Euclidean section}
%%%%%%%%%%%%%%%%%%%%%%%%%%%%%%%%

One of the key sections of the \EX spacetime is the Euclidean section. The associated metric is obtained from Eq.~(\ref{guimetric})  by assuming that
$(\sigma, \xi, y,z)\in{\Bbb R^4}$, where $\sigma\equiv-i\eta$. A straightforward substitution leads to
\begin{equation} \mlab{euclidmetric}
ds^2 \, = \, \frac{d\sigma^2 \, + \, d\xi^2}{\alpha^2\left(\sigma^2 \, + \, \xi^2\right)}
 \, + \, dy^2 \, + \, dz^2\, .
\end{equation}
By use of the transformations
\begin{eqnarray} \mlab{euclidTrans}
%\left\{
%\begin{array}{rcl}
\sigma &\hspace{-1mm} =\hspace{-1mm}&  (1/\alpha)\,
\exp\left(\alpha x\right) \sin\left(\alpha\tau\right),  \\[2mm]
\xi &\hspace{-1mm} =\hspace{-1mm}& (1/\alpha)\,
\exp\left(\alpha x\right) \cos\left(\alpha\tau\right),
%\end{array}
%\right.
\label{euclidTrans2}
\end{eqnarray}
the metric becomes that of a cylindrical (Euclidean) spacetime, i.e.
\begin{equation}
ds^2 \, = \, d\tau^2 \, + \, dx^2 \, + \, dy^2 \, + \, dz^2\, ,
\end{equation}
where the temporal direction $\alpha\tau$  is $2\pi$-periodic. In our following considerations,
we will restrict the basic temporal interval to $0 \leq \tau \leq \beta$.
On the QFT level, this setting together with the single-valuedness of quantum fields
will yield the typical periodicity property of the Euclidean propagator, cf. Section~4.
%a periodic structure, namely $\tau\equiv\tau
%+\beta$.

%In the $\sigma-\xi$ plane, $\tau$ plays the role of polar angle.

%%%%%%%%%%%%%%%%%%%%%%%%%%%%%%%%%%%%%%%%%%%%%%%%%%%%%%%%
\subsection{Lorentzian section} \label{subsecAC}
%%%%%%%%%%%%%%%%%%%%%%%%%%%%%%%%%%%%%%%%%%%%%%%%%%%%%%%
%
\begin{figure}[t]
\centering
\includegraphics[scale=0.4]{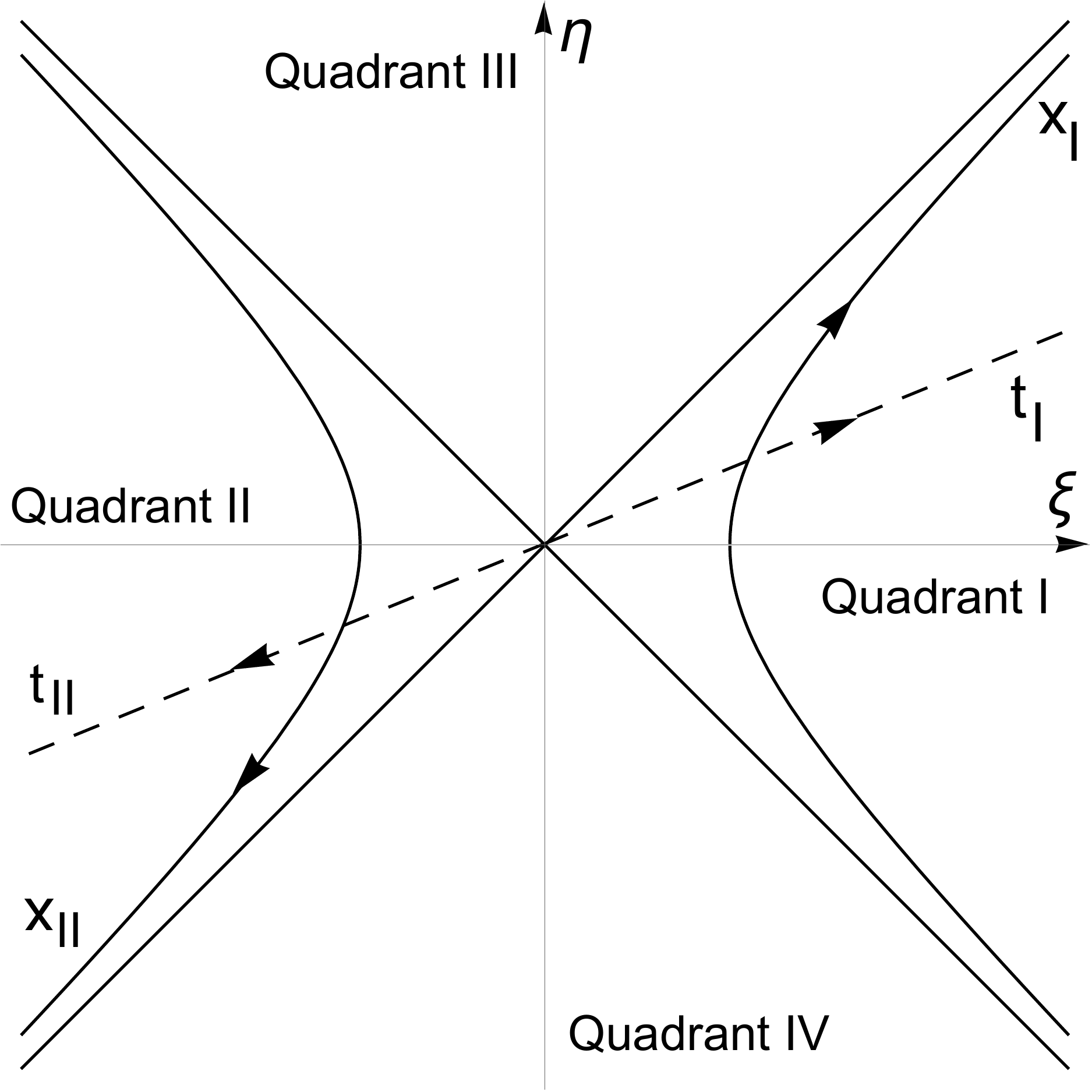}
\caption{Lorentzian section of \EX spacetime: the solid lines
represent the singularities at $\xi^2-\eta^2=0$. On the (dashed) straight lines
time is constant, while on the hyperbolas the Minkowski coordinate $x$
is constant.
Note that times $t_\I$ and $t_\II$ flow in opposite directions as a consequence of opposite orientations
of ensuing time-like Killing vectors in regions $R_\I$ and $R_\II$.}
\label{fig2}
\end{figure}
The second important section is the Lorentzian section. In this case, the line element is given by Eq.~(\ref{guimetric})
with $(\eta, \xi, y,z)\in{\Bbb R^4}$.  The ensuing metric is singular on the
two hyperplanes $\eta=\pm\,\xi$, which we will call
``event-horizons''~\cite{note}. These divide \EX spacetime into four regions
denoted by $R_\I, R_\II,R_\III$ and $R_\IV$ (see Fig.~\ref{fig2}).

In the first two regions, one can define two sets of tortoise-like
coordinates $x_\IQ^{\mu}\in\R^4$ by $x_\IQ^{\mu}=(t_\IQ,x_\IQ,y,z)$ where
\begin{eqnarray}
&\mbox{in $R_\I$:} & \quad \left\{
\begin{array}{rcl}
\eta & \hspace{-1mm}=&\hspace{-1mm} +(1/\alpha)\,
\exp\left(\alpha x_\I\right)\sinh\left(\alpha t_\I\right), \\[2mm]
\xi &\hspace{-1mm}=&\hspace{-1mm} +(1/\alpha)\,
\exp\left(\alpha x_\I\right)\cosh\left(\alpha t_\I\right),
\end{array}\right.
\label{TransI}\\[3mm]
&\mbox{in $R_\II$:} & \quad \left\{
\begin{array}{rcl}
\eta &\hspace{-1mm}=&\hspace{-1mm} - (1/\alpha)\,
\exp\left(\alpha x_\II\right)\sinh\left(\alpha t_\II\right), \\[2mm]
\xi &\hspace{-1mm}=&\hspace{-1mm} - (1/\alpha)\,
\exp\left(\alpha x_\II\right) \cosh\left(\alpha t_\II\right).
\end{array}
\right.
\label{TransII}
\end{eqnarray}

Similar transformations hold also in regions $R_\III$ and $R_\IV$
(see Refs.~\cite{GUI90,FV,FV1}) but for our purposes it will be sufficient to consider only the first two regions.  In terms of the new coordinates $x_\IQ^{\mu}$, the metric in Eq.~(\ref{guimetric}) becomes the usual Minkowski one with the metric signature in the spacelike convention
\begin{equation}
ds^2 \, = \, -dt^2_\IQ \, + \, dx_\IQ^2 \, + \, dy^2 \, + \, dz^2\, ,
\end{equation}
(and similarly for $R_\III$ and $R_\IV$). It thus arises that regions $R_\I$ to $R_\IV$ represent four copies of the Minkowski spacetime glued
together along the ``event-horizons''~\cite{note}, making up together the Lorentzian section of \EX spacetime.

Although Eqs.~(\ref{TransI})-(\ref{TransII}) formally correspond to Rindler transformations~\cite{Rin}, the $(t_{\I,\II},x_{\I,\II},y,z)$
coordinates should not be confused with the Rindler ones, since in those coordinates the metric takes the standard Minkowski form.
Indeed, an observer whose world-line is the hyperbola described for example by $(x_\I(t_\I),y(t_\I),z(t_\I))=(x_0,y_0,z_0)$ moves in the Lorentzian wedge $R_\I$ inertially and thus cannot be identified with an accelerated observer as in the Rindler case. The r\^{o}le of the inertial and non-inertial coordinates in Eqs.~(\ref{TransI})-(\ref{TransII}) are actually reversed with respect to the Rindler case~\cite{Rin,GUI90}.

We now study the analytic properties of the transformations given in
Eqs.~(\ref{TransI})-(\ref{TransII}). In the null coordinates $\xi^\pm=\eta\pm\xi$, we can
rewrite these equations as
\begin{eqnarray}
&\mbox{in $R_\I$:}& \;
\; \left\{\begin{array}{rcl}
\xi^+_>(t_\I,x_\I) &\hspace{-1.5mm}=\hspace{-1.5mm}&
+(1/\alpha)\exp\left[+\alpha\left(t_\I\,+\,x_\I\right)\right],
\\[2mm]
\xi^-_<(t_\I,x_\I) &\hspace{-1.5mm}=\hspace{-1.5mm}&
-(1/\alpha)\exp\left[-\alpha\left(t_\I\,-\,x_\I\right)\right],
\end{array}\right.
\\[2mm]
&\hspace{0.5mm}\mbox{in $R_\II$:}& \; \; \left\{\begin{array}{rcl}
\xi^+_<(t_\II,x_\II) &\hspace{-1.5mm}=\hspace{-1.5mm}&
-(1/\alpha)\exp\left[+\alpha\left(t_\II\,+\,x_\II\right)\right],
\\[2mm]
\xi^-_>(t_\II,x_\II) &\hspace{-1.5mm}=\hspace{-1.5mm}&
+(1/\alpha)\exp\left[-\alpha\left(t_\II\,-\,x_\II\right)\right],
\end{array}\right.
\end{eqnarray}
where  the subscripts $<$ and $>$ have been added to the variables
$\xi^\pm$ to indicate their ranges, i.e.~one has $\xi^\pm_>>0$ and
$\xi^\pm_<<0$. The reciprocals of these transformations are
\begin{eqnarray}
&\mbox{in $R_\I$:}& \;
\; \left\{\begin{array}{rcl}
t_\I(\xi^+_>,\xi^-_<) &\hspace{-1.5mm}=\hspace{-1.5mm}& \displaystyle
\frac{1}{2\alpha}\ln\left(\displaystyle-\frac{\xi^+_>}{\xi^-_<}\right),
\\[4mm]
x_\I(\xi^+_>,\xi^-_<) &\hspace{-1.5mm}=\hspace{-1.5mm}&\displaystyle
\frac{1}{2\alpha}\ln\left(-\alpha^2\xi^+_>\xi^-_<\right),
\end{array}\right.
\label{RecI} \\[2mm]
&\mbox{in $R_\II$:}& \; \;\left\{\begin{array}{rcl}
t_\II(\xi^+_<,\xi^-_>) &\hspace{-1.5mm}=\hspace{-1.5mm}& \displaystyle
\frac{1}{2\alpha}\ln\left(\displaystyle-\frac{\xi^+_<}{\xi^-_>}\right),
\\[4mm]
x_\II(\xi^+_<,\xi^-_>) &\hspace{-1.5mm}=\hspace{-1.5mm}&\displaystyle
\frac{1}{2\alpha}\ln\left(-\alpha^2\xi^+_<\xi^-_>\right).
\end{array}\right.
\label{RecII}
\end{eqnarray}
Equations (\ref{RecI}) and (\ref{RecII}) are defined in regions $R_\I$
and $R_\II$, respectively. We now would like to analytically continue
these expressions to obtain the functions $t_\I(\xi)$, $t_\II(\xi)$
and $x_\I(\xi)$, $x_\II(\xi)$ defined in $R_\I\cup R_\II$. This
amounts to extend these expressions from positive or negatives values
of $\xi^\pm$ to their negative or positive values respectively.  In
order to do this, we choose to perform the analytic extension
in the {\it lower} half-planes of both the $\xi^+$ and $\xi^-$
complex planes for reasons which will become clear below. In other
words, we assume that $-\pi<\arg\xi^\pm\leq\pi$, or equivalently that
the cuts in the $\xi^\pm$ complex planes are given by $\R_-+i0_+$.
\begin{figure}[t]
\begin{center}
\includegraphics[scale=0.3]{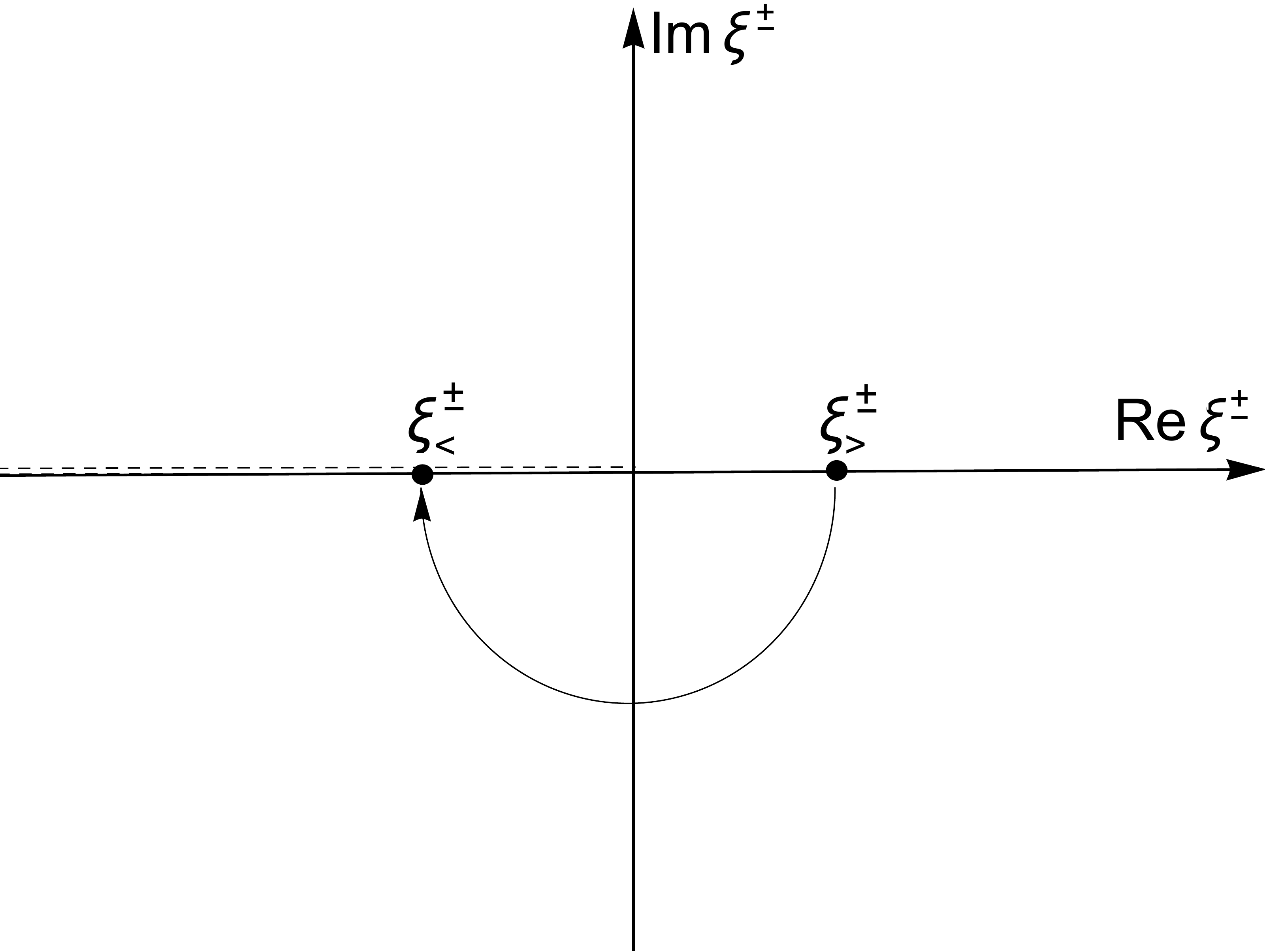}
\end{center}
\caption{Analytic extension in the \emph{lower} half-planes of the $\xi^+$
and $\xi^-$ complex planes. The dashed line $\R_-+i0_+$ is the branch cut of the logarithm
in Eq.~(\ref{analcont1}).
 }
\label{fig3}
\end{figure}
It is not possible to perform the analytic continuation with respect to
the two variables $\xi^\pm$ {\it at once}, otherwise an erroneous
result would be obtained. To fix the ideas, we choose to perform the
extension first in the $\xi^+$ variable and then in $\xi^-$ (the
choice of the opposite order gives the same result for our particular
purposes).  If $\xi^\pm_<$ is the analytic continuation of $\xi^\pm_>$
from positive to negative values, one has (see Fig.~\ref{fig3})
\begin{eqnarray}
\label{analcont1}
\ln\left(-\xi^\pm_<\right) &\hspace{-1.5mm}=\hspace{-1.5mm}& \ln\left(+\xi^\pm_>\right)\,+\,i\pi\, ,
\\[2mm]
\ln\left(+\xi^\pm_>\right) &\hspace{-1.5mm}=\hspace{-1.5mm}& \ln\left(-\xi^\pm_<\right)\,-\,i\pi\, .
\label{analcont2}
\end{eqnarray}
This implies
\begin{eqnarray}
\hspace{5.5mm}\ln\left(-\frac{\xi^+_<}{\xi^-_>}\right) &\hspace{-1.5mm}=\hspace{-1.5mm}&
\ln\left(-\frac{\xi^+_>}{\xi^-_<}\right) \,+\, i2\pi\, ,
\\[2mm]
\ln\left(-\alpha^2\xi^+_<\xi^-_>\right) &\hspace{-1.5mm}=\hspace{-1.5mm}&
\ln\left(-\alpha^2\xi^+_>\xi^-_<\right),
\end{eqnarray}
which means that these expressions are the analytic continuations of
each other. Consequently, by using Eqs.~(\ref{RecI})-(\ref{RecII}), one obtains
in $R_\I\cup R_\II$
\begin{eqnarray}
\label{ACt}
t_\II(\xi)  &\hspace{-1.5mm}=\hspace{-1.5mm}&  t_\I(\xi) \, + \, i\,\beta/2, \\[2mm]
x_\II(\xi) &\hspace{-1mm} =\hspace{-1mm}& x_\I(\xi)\, .
\label{ACtx}
\end{eqnarray}

We note in passing that we can call the hypersurfaces $\xi^2 \,-\, \eta^2 =0$
``horizons'' only in the sense that an inertial observer in
region $R_{\I}$ cannot receive any signal sent from $\xi^- \,=\,  0$, and
cannot send any signal to $\xi^+ \,=\,  0$. So the hypersurface
$\xi^-  \,=\, 0$ or $\xi^+ \,=\, 0$ can formally be called a ``future horizon''
$\mathcal{H}^+$ or ``past horizon'' $\mathcal{H}^-$, respectively for an inertial
observer in region $R_{\I}$. Analogous conclusion holds, of course also for an inertial observer 
in region $R_{\II}$.  
%although they are not the horizons in the usual
%sense.

%%%%%%%%%%%%%%%%%%%%%%%%%%%%%%%%%%%%%%%%%%%%%%%%%%%%%%%%%
\subsection{Extended Lorentzian section} \label{sub:ELS}
%%%%%%%%%%%%%%%%%%%%%%%%%%%%%%%%%%%%%%%%%%%%%%%%%%%%%%%%%%

Let us now consider a class of complex sections of \EX spacetime generated
from the Lorentzian section by shifting the Minkowski time coordinate
in the imaginary direction but {\it only} in the region $R_\II$, namely
\begin{eqnarray}\mlab{shift}
\begin{array}{lll}
&\mbox{in $R_\I\cup R_\III\cup R_\IV$:}
\qquad & t_q \rightarrow {t_q}_{_\delta}\,=\,t_q, \;
\;\;\; q\, =\, I,II \mbox{~and~} III\, ,
\\[2mm] &\mbox{in $R_\II$:} \qquad & t_\II
\rightarrow t_{\IId}\,=\, t_\II\, +\, i\beta\delta\, ,
\end{array}
\label{delta}
\end{eqnarray}
where $\delta\in[-1/2,1/2]$. The reason for this interval will become clear shortly.
We shall call this one-parametric class of sections as ``extended
Lorentzian section'' and denote by $R_\IId$ the image of the region
$R_\II$ resulting from this shift. In $R_\IId$ the \EX coordinates become
complex variables and are transformed according to $(\eta,\xi)
\rightarrow (\etad,\xid)$ where, from Eq.~(\ref{shift}), we have
\begin{eqnarray}\mlab{mex2a}
{\mbox{in}}~R_\IId: \;\;\; \left\{
\begin{array}{rcl}
\etad &\hspace{-1.5mm}=\hspace{-1.5mm}& -(1/\alpha)\,\exp\left(\alpha x_\II\right)
\sinh\left[\alpha\left(t_\II\,+\,i\beta\delta\right)\right],
\\[2mm]
\xid &\hspace{-1.5mm}=\hspace{-1.5mm}& -(1/\alpha)\,\exp\left(\alpha x_\II\right)
\cosh\left[\alpha\left(t_\II\,+\, i \beta \delta\right)\right].
\end{array}
\right.
\end{eqnarray}
In terms of the real \EX variables, one can write
\begin{eqnarray}
\label{mex31}
\etad &\hspace{-1.5mm}=\hspace{-1.5mm}& +\eta \, \cos\left(2\pi\delta\right)
\,+\, i\xi\,\sin\left(2\pi\delta\right), \\[2mm]
i\xid &\hspace{-1.5mm}=\hspace{-1.5mm}& -\eta \, \sin\left(2\pi\delta\right)
\,+\, i\xi\,\cos\left(2\pi\delta\right),
\mlab{mex3}
\end{eqnarray}
or, equivalently, in terms of null coordinates
\begin{eqnarray}\mlab{mex3b}
%\left\{ \begin{array}{rcl}
\xid^+ &\hspace{-1.5mm}=\hspace{-1.5mm}& \exp\left(+i2\pi\delta\right)\,\xi^+, \\[2mm]
\xid^- &\hspace{-1.5mm}=\hspace{-1.5mm}& \exp\left(-i2\pi\delta\right)\,\xi^-.
%\end{array}
%\right.
\label{defxipm}
\end{eqnarray}
The time shift in Eq.~(\ref{shift}) thus induces a rotation in the $(\eta,i\xi)$ plane of
$R_\II$.  By using the rotated coordinates, the metric can then be recast into the form
\begin{equation}\mlab{mex1}
ds^2 \, = \, \frac{-d\eta_\delta^2 \, + \, d\xi_\delta^2}
{\alpha^2\left(\xi_\delta^2 \, - \, \eta^2_\delta\right)} \,+ \, dy^2 \, + \, dz^2\, ,
\end{equation}
and is thus unchanged by the time shift Eq.~(\ref{shift}), which is therefore an isometry
of the \EX spacetime. After the time shift, Eqs.~(\ref{ACt})-(\ref{ACtx}) become 
\begin{eqnarray}
\label{ACtxd1}
t_\IId(\xid) &\hspace{-1.5mm}=\hspace{-1.5mm}&\ \!
 t_\I(\xid) \,+\, i\,\beta\left(1/2 \,+\, \delta\right), 
\\[2mm]
x_\IId(\xid) & \hspace{-1mm}=\hspace{-1mm}&\ \! x_\I(\xid)\, .
%\end{array} \right.
\label{ACtxd}
\end{eqnarray}

Let us finally comment on Killing fields in
%both Euclidean and
(extended) Lorentzian section.
Since a timelike Killing vector defines a preferred time coordinate in a time-independent spacetime~\cite{Straumann}, we can expect
(with some foresight) that its structure is pertinent  for the understanding of the connection between the (extended) Lorentzian section and the POM.
It will also prove useful when we will discus TFD in Sec.~\ref{Sec.4.3}.
%
%in order to
%would be particularly relevant in TFD and POM.
%

In order to find the Killing vector field $\kappa$ in various sections of \EX spacetime, we need to solve the Killing equation
$({\mathcal{L}}_{\kappa} \ \! g)_{\mu\nu} = \kappa^{\lambda}g_{\mu\nu, \lambda} + g_{\lambda\nu} \kappa^{\lambda}_{\ \! , \mu}
+ g_{\mu \lambda} \kappa^{\lambda}_{\ \! , \nu}
%= \kappa_{\mu;\nu}  + \kappa_{\nu;\mu}
= 0$. Here ${\mathcal{L}}_{\kappa}$ is the Lie derivative along the vector field $\kappa$ and $g_{\mu\nu}$ is the pullback metric on a given section. For instance,
in the Lorentzian section we have
\begin{equation}
\kappa \, = \, \alpha\left(\xi \ \! \frac{\partial}{\partial \eta} \, + \, \eta \ \! \frac{\partial}{\partial \xi} \right),
\label{Lorentz.case.a}
\end{equation}
which is clearly timelike in $R_\I \cup R_\IId$ as there $\kappa^2 = g_{\mu \nu} \kappa^{\mu} \kappa^{\nu} = -1$. The parameter $\alpha$ was introduced in Eq.~(\ref{Lorentz.case.a}) so that  the components of $\kappa$ become dimensionless and normalized to $-1$. Along the same lines, we see that, in the extended Lorentzian section, we obtain from Eq.~(\ref{mex1}) 
\begin{equation}
\kappa \, = \, \alpha\left(\xi_{\delta} \ \! \frac{\partial}{\partial \eta_{\delta}} \, + \, \eta_{\delta} \ \! \frac{\partial}{\partial \xi_{\delta}} \right).
\label{Lorentz.case.b}
\end{equation}
Again, $\kappa$ is timelike in $R_\I \cup R_\IId$. The integral curves of the above Killing vector fields satisfy  equations ${d \xi_{\delta}}/{d s} \,=\, \alpha \eta_{\delta}$ and ${d \eta_{\delta}}/{d s} \,=\, \alpha \xi_{\delta}$, which yield a parametric representation of orbits in the form
\begin{eqnarray}
\label{orbits1}
\xi_{\delta}(s) &\hspace{-1.5mm}=\hspace{-1.5mm}& \ \! \xi_{\delta}^0 \ \! \cosh[\alpha(s \,+\, i\beta \delta)]\, ,  \\[2mm]
\eta_{\delta}(s) &\hspace{-1.5mm}=\hspace{-1.5mm}& \ \! \eta_{\delta}^0 \ \! \sinh[\alpha(s \,+\, i\beta \delta)]\, ,
\label{orbits}
\end{eqnarray}
where $\delta$ is the shift parameter introduced in Eq.~(\ref{delta}), in particular, $\delta \,=\, 0$ in $R_\I$. 
The integration constants $\xi_{\delta}^0$ and $\eta_{\delta}^0$ depend on an actual position $x_{\I}$ or $x_{\II}$ of the observer.  
Equations~(\ref{orbits1})-(\ref{orbits}) precisely coincide with the world-lines of a static observer in respective Minkowski wedges. Note that the Killing 
vectors in regions $R_\I$ and $R_\IId$  have mutually opposite orientations of their real parts.

Analogous reasonings yield in the Euclidean section the Killing vector
\begin{equation}
\kappa \, = \, \alpha\left(\xi \ \! \frac{\partial}{\partial \sigma} \, - \, \sigma \ \! \frac{\partial}{\partial \xi} \right),
\label{Eucl.cas.a}
\end{equation}
which is the Euclidean analogue of the timelike Killing vector.
The integral curves of the above $\kappa$ are circles around the origin with radius $R = e^{\alpha x}/\alpha$ (here $R^2 = \sigma^2 + \xi^2$). 
Hence these orbits can be identified with preferred inverse-temperature coordinate in the Euclidean section.

Let us note, finally, that the coordinates $x_\IQ^{\mu}=(t_\IQ,x_\IQ,y,z)$, $x_{\I,\IId}^{\mu}=(t_{\I,\IId},x_{\I,\IId},y,z)$ and $x^{\mu} = (\tau,x,y,z)$ 
are adapted to the Killing fields in Eqs.~(\ref{Lorentz.case.a}), (\ref{Lorentz.case.b}) and (\ref{Eucl.cas.a}), respectively.

%%%%%%%%%%%%%%%%%%%%%%%%%%%%%%%%%%%%%%%%%%%%%%
\section{Field quantization in \EX spacetime}
%%%%%%%%%%%%%%%%%%%%%%%%%%%%%%%%%%%%%%%%%%%%%%
%
%To put some meat on the bare bones, we discuss our point with
%scalar quantum fields.
For sake of simplicity, let us now consider a free scalar field with a support in \EX spacetime.
The corresponding generalization to the fermionic sector is quite straightforward and one may, for instance,
follow the line of reasonings presented in Ref.~\cite{GUI92ferm}.

We denote the ``global'' scalar field in \EX coordinates as $\Phi(\xi)$. It satisfies
the Klein--Gordon equation
\begin{equation} \mlab{kge}
(\Box \,-\, m^2) \Phi(\xi) \, = \,  0\, ,
\end{equation}
where $\Box \,=\, {g}^{-1/2}\partial_\mu\hspace{0.2mm} g^{\mu\nu}{g}^{-1/2}\partial_\nu$ is the Laplace--Beltrami operator, with $g(\xi) \,=\, \left|\mathrm{det}\hspace{0.4mm} g_{\mu\nu}\right|$. We define the inner product of two Klein--Gordon fields by
\begin{equation}
(\Phi_1,\Phi_2) \, = \, -i\int_\Sigma\hspace{-1mm} d\Sigma\, {\left(g(\xi)\right)}^{1/2}\hspace{0.1mm}
\Phi_1(\xi)\,n^\nu\stackrel{\leftrightarrow}{\partial}_\nu \Phi_2^*(\xi)\, ,
\label{scalarproduct}
\end{equation}
where the integral is taken over a Cauchy hypersurface $\Sigma$ and $n^\nu$ an orthonormal
vector to this hypersurface.

%%%%%%%%%%%%%%%%%%%%%%%%%%%%%%%%
\subsection{Euclidean section}
%%%%%%%%%%%%%%%%%%%%%%%%%%%%%%%%

In the Euclidean section of \EX spacetime, one has $\Phi=\Phi(\sigma,\xi,y,z)$. The
field in the $\tau$-$x$ coordinates in Eqs.~(\ref{euclidTrans})-(\ref{euclidTrans2})
shall be denoted by $\phi=\phi(\tau,x,y,z)$. These two fields are
related by
\begin{equation}
\phi(\tau,x,y,z) \, = \, \Phi(\sigma(\tau,x),\xi(\tau,x),y,z)\, .
\end{equation}
Because of the periodic nature of the time $\tau$ and presumed single-valuedness of the field, the following condition must be satisfied
\begin{equation}
\phi(\tau,x,y,z) \, = \, \phi(\tau+\beta,x,y,z)\, .
\label{periodicfield}
\end{equation}
This periodic boundary condition will prove to be important in Section~4.
Note that Eq.~(\ref{periodicfield}) is nothing but the familiar Kubo--Martin--Schwinger (KMS) boundary condition for Euclidean fields~\cite{LW87,BJV}.

%%%%%%%%%%%%%%%%%%%%%%%%%%%%%%%%%%%%%%%%%%%%%%%%%%%%%%%%%%%
\subsection{Lorentzian section} \label{Lorentzian section}
%%%%%%%%%%%%%%%%%%%%%%%%%%%%%%%%%%%%%%%%%%%%%%%%%%%%%%%%%%%

Lorentzian section, as we have seen, is made up of four different
regions, each of them being a complete Minkowski spacetime.  Since our primary
interest is only in regions $R_\I$ and $R_\II$, we shall limit ourselves to
consider the quantum field over these two regions.  Our aim is to find an
expansion for the global field $\Phi$ in the joining $R_\I \cup
R_\II$.

Let us start by defining the ``local'' fields $\phi^\I(x_\I)$ and
$\phi^\II(x_\II)$ by
\begin{eqnarray} \label{defphi}
\Phi(\xi) \ = \ \left\{
\begin{array}{lll}
\phi^\I(x_\I(\xi)), &\quad& \mbox{when $\xi\in R_\I$}\, , \\[2mm]
\phi^\II(x_\II(\xi)), && \mbox{when $\xi\in R_\II$}\, .
\end{array}\right.
\end{eqnarray}
They have support in $R_\I$ and $R_\II$, respectively.
By choosing $\Sigma$ to be any of the one-parametric class of hypersurfaces $\eta = a\xi$ (with $-1<a<1$), 
we obtain from Eq.~(\ref{scalarproduct}) that the global inner product assumes the form
\begin{equation}
(\Phi_1,\Phi_2) \, = \, \langle \phi^\I_1,\phi^\I_2 \rangle \, + \, \langle \phi^\II_1,\phi^\II_2 \rangle \, ,
\label{SP}
\end{equation}
where $\langle \,,\, \rangle$ is the local inner product in Minkowski spacetime
\begin{equation}
\langle \phi_1,\phi_2 \rangle \, = \, -i\int_{\R^3}\hspace{-1.9mm} d^3x\,
\phi_1(x)\stackrel{\leftrightarrow}{\partial}_t\phi_2^*(x)\, .
\label{SPM}
\end{equation}
This is nothing but the usual Klein--Gordon inner product known from relativistic quantum theory~\cite{Itzykson}.

In \EX spacetime covered by the $x_\IQ^{\mu}$ coordinates defined by
Eqs.~(\ref{TransI})-(\ref{TransII}), the solutions of the
Klein--Gordon equation are just plane waves restricted to a given
region. So, we can write explicitly
\begin{eqnarray}
u_\Vk(x_\I) &\hspace{-1.5mm}=\hspace{-1.5mm}& \left(4\pi\omega_\Vk\right)^{-\frac{1}{2}}\,
e^{i\left(-\omega_\Vk\,t_\I\,+\,\Vk\cdot\Vx_\I\right)},
\mlab{MMu} \\[2mm]
v_\Vk(x_\II) &\hspace{-1.5mm}=\hspace{-1.5mm}& \left(4\pi\omega_\Vk\right)^{-\frac{1}{2}}\,
e^{i\left(+\omega_\Vk\,t_\II\,+\,\Vk\cdot\Vx_\II\right)},
\mlab{MMv}
\end{eqnarray}
where $\omega_\Vk= \sqrt{{|\Vk|}^2+m^2}$. Starting from these Minkowski modes,
one defines the two wave functions $U_\Vk(\xi)$ and $V_\Vk(\xi)$ with
support in $R_\I$ and $R_\II$, respectively, by
\begin{eqnarray}
U_\Vk(\xi) &=&  \left\{
\begin{array}{lcl}
u_\Vk(x_\I(\xi)), &\quad&
\mbox{when $\xi\in R_\I$}, \\[2mm]
0, && \mbox{when $\xi\in R_\II$},
\end{array}\right.
\label{defU}\\[2mm]
V_\Vk(\xi) &=&
\left\{\begin{array}{lcl}
0, && \mbox{when $\xi\in R_\I$}, \\[2mm]
v_\Vk(x_\II(\xi)), &\quad& \mbox{when $\xi\in R_\II$}.
\end{array}\right.
\label{defV}
\end{eqnarray}
Their power spectra with respect to the momenta conjugated to $\xi^+$
and $\xi^-$ contain negative contributions, which are furthermore not
bounded from below. The sets of functions $\left\{
U_\Vk(\xi), U^*_{-\Vk}(\xi)\right\}_{\Vk\in\R^3}$ and $\left\{
V_\Vk(\xi), V^*_{-\Vk}(\xi)\right\}_{\Vk\in\R^3}$ defined on $R_\I$
and $R_\II$, respectively, are thus over-complete, since the same energy
contribution (i.e.~momentum contribution conjugate to $\eta$) can
appear twice in these sets. In other words, the energy spectra of
$U_\Vk$ and $U^*_{-\Vk}$ overlap, and so do the spectra of $V_\Vk$ and
$V^*_{-\Vk}$. Therefore, these sets cannot be used as a basis in their
respective supports and their joining is clearly not a
basis in $R_\I\cup R_\II$.
%\smallskip
In order to construct a basis in $R_\I\cup R_\II$, we could solve, for instance, the
Klein--Gordon equation in \EX coordinates to obtain the field modes in
these coordinates. The Bogoliubov transformations resulting
from this choice of basis would be, however, quite complicated. So instead of following this route,
we shall construct basis elements with positive energy
spectra from the wave functions $u_\Vk(x_\I(\xi))$
and $v^*_{-\Vk}(x_\II(\xi))$. We will further demand these basis elements should be analytic functions in
the lower complex planes of $\xi^+$ and $\xi^-$, so that their
spectra with respect to the momenta conjugate to $\xi^+$ and $\xi^-$ contain only positive contributions.
Consequently, they will have positive energy spectra.

To this end, we analytically extend the two wave functions $u_\Vk(x_\I(\xi))$
and $v^*_{-\Vk}(x_\II(\xi))$ in the lower complex planes of $\xi^+$
and $\xi^-$ (as in Section \ref{subsecAC}, the cut in the complex
planes is represented by $\R_-+i0_+$). By applying Eqs.~(\ref{ACt})-(\ref{ACtx}),
we obtain directly
\begin{eqnarray}
u_\Vk(x_\I(\xi)) &\hspace{-1.5mm}=\hspace{-1.5mm}& e^{-\frac{\beta}{2}\omega_\Vk}\,v^*_{-\Vk}(x_\II(\xi))\, ,
\label{ACuv} \\[2mm]
v_\Vk(x_\II(\xi)) &\hspace{-1.5mm}=\hspace{-1.5mm}& e^{-\frac{\beta}{2}\omega_\Vk}\,u^*_{-\Vk}(x_\I(\xi))\, .
\label{ACvu}
\end{eqnarray}
The expressions on the right and left hand sides of these
equations are analytic continuations of each other. We are therefore led
to introduce the following two normalized linear combinations
\begin{eqnarray}
\label{globalmodes1}
F_{\Vk}(\xi) &\hspace{-1.5mm}=\hspace{-1.5mm}&\left(1-f_\Vk\right)^{-\frac{1}{2}}
\left[ U_\Vk(\xi) \,+\, f_\Vk^{\frac{1}{2}}\,V^*_{-\Vk}(\xi)\right],
\\[3mm]
{\widetilde F}_{\Vk}(\xi) &\hspace{-1.5mm}=\hspace{-1.5mm}& \left(1-f_\Vk\right)^{-\frac{1}{2}}
\left[ V_{\Vk}(\xi) \,+\, f_\Vk^{\frac{1}{2}}\,U^*_{-\Vk}(\xi)\right],
\label{globalmodes}
\end{eqnarray}
where $f_\Vk=e^{-\beta\omega_\Vk}$ and $U_\Vk(\xi)$ and $V_\Vk(\xi)$ are defined in Eqs.~(\ref{defU}) and (\ref{defV}), respectively.
These global wave functions are still solutions of the Klein--Gordon equation. Moreover, they
are analytic in $R_\I\cup R_\II$ and in particular at the origin
$\xi^+=\xi^-=0$.  Since they are analytic complex functions in the
lower complex planes of $\xi^+$ and $\xi^-$, their energy spectra are positive. The set $\{F_{\Vk}, F^*_{-\Vk},
{\widetilde F}_\Vk, {\widetilde F}^*_{-\Vk} \}_{\Vk\in\R^3}$ is thus
complete (but not over-complete) over the joining $R_\I\cup
R_\II$. Furthermore it is an orthogonal set since
\begin{eqnarray}
(F_\Vk,F_\Vp)& \hspace{-1.3mm}=\hspace{-1.3mm} & (\widetilde{F}^*_\Vk,\widetilde{F}^*_\Vp)
\,=\,  +\,\delta^3(\Vk-\Vp)\, ,
\\[2mm]
(F^*_\Vk,F^*_\Vp)& \hspace{-1.3mm}=\hspace{-1.3mm} &(\widetilde{F}_\Vk,\widetilde{F}_\Vp)
\,=\,  -\delta^3(\Vk-\Vp)\, ,
\end{eqnarray}
with all the other scalar products vanishing.

On one hand, we can expand the local scalar fields in terms of the Minkowski modes given
in Eqs.~(\ref{MMu})-(\ref{MMv}) as follows
\begin{eqnarray}
\phi^\I(x_\I) &\hspace{-1.5mm}=\hspace{-1.5mm}&\int\hspace{-1mm} d^3k \left[ a^\I_\Vk\,u_\Vk(x_\I)\,+\,
a^{\I\dagger}_\Vk\,u^*_\Vk(x_\I)\right],
\label{ExpI}\\[1mm]
\phi^\II(x_\II) &\hspace{-1.5mm}=\hspace{-1.5mm}& \int\hspace{-1mm} d^3k \left[ a^\II_\Vk\,v_\Vk(x_\II)\,+\,
a^{_\II\dagger}_\Vk\,v^*_\Vk(x_\II)\right].
\label{ExpII}
\end{eqnarray}
On the other hand, the global scalar field can be expanded in terms of the $F$-modes given in Eqs.~(\ref{globalmodes1})-(\ref{globalmodes}) as
\begin{equation}
{ \Phi(\xi)\,=\, \int\hspace{-1mm} d^3k \Big[
b_\Vk\,F_\Vk(\xi) \,+\, b^\dagger_\Vk\,F^*_{\Vk}(\xi)
 }\, +\,\tilde{b}_\Vk\,\widetilde{F}_{-\Vk}(\xi)
\,+\,\tilde{b}_\Vk^\dagger\,\widetilde{F}^*_{-\Vk}(\xi) \Big]\hspace{-0.3mm}.
\mlab{Exp}
\end{equation}
These three expansions define the local and global creation and
annihilation operators, which are connected to each other by Bogoliubov
transformations. In order to obtain these transformations, we use the definition Eq.~
(\ref{defphi}) relating the local and global fields and the
field expansions Eqs.~(\ref{ExpI}), (\ref{ExpII}) and (\ref{Exp}). A straightforward calculation
leads to
\begin{eqnarray}\label{bogol}
b_\Vk &\hspace{-1mm} =\hspace{-1mm} &  a^\I_\Vk\,\cosh\theta_\Vk\, -\, a^{\II\dagger}_{-\Vk}\,\sinh\theta_\Vk\, ,
\\[2mm]
{\tilde b}_\Vk &\hspace{-1mm} =\hspace{-1mm} &
a^\II_{-\Vk}\,\cosh\theta_\Vk\,-\,a^{\I\dagger}_\Vk\,\sinh\theta_\Vk\, ,
\label{bogol2}
\end{eqnarray}
where $\sinh^2\theta_\Vk = n(\omega_\Vk)=(e^{\beta\omega_\Vk } -1)^{-1}$ is the Bose--Einstein distribution.

%%%%%%%%%%%%%%%%%%%%%%%%%%%%%%%%%%%%%%%%%%%
\subsection{Extended Lorentzian section}
%%%%%%%%%%%%%%%%%%%%%%%%%%%%%%%%%%%%%%%%%%%%

By following the above outlined procedure, we can now construct a set of
positive energy modes in the extended Lorentzian section
introduced in Section~\ref{sub:ELS}. We start by considering the region
$R_\II$, where the plane-wave set
$\{v_\Vk(x_\II),v^*_{-\Vk}(x_\II)\}_{\Vk\in\R^3}$ has the form
\begin{eqnarray}
v_\Vk(x_\II) &\hspace{-1.5mm}=\hspace{-1.5mm}& \left(4\pi\omega_\Vk\right)^{-\frac{1}{2}}\,
e^{i\left(+\omega_\Vk\,t_\II\,+\,\Vk\cdot\Vx_\II\right)},
\\[2mm]
v_{-\Vk}^*(x_\II) &\hspace{-1.5mm}=\hspace{-1.5mm}& \left(4\pi\omega_\Vk\right)^{-\frac{1}{2}}\,
e^{i\left(-\omega_\Vk\,t_\II\,+\,\Vk\cdot\Vx_\II\right)}.
\end{eqnarray}
Under the time shift Eq.~(\ref{shift}), this set is transformed into
$\{v_\Vk(x_{\IId}),v^\sharp_{-\Vk}(x_{\IId})\}_{\Vk\in\R^3}$, where the symbol
 $\ast$ has been replaced by $\sharp$ because
$v^\sharp_{-\Vk}(x_{\IId})$ is no longer the complex conjugate of
$v_\Vk(x_{\IId})$. In fact, one has
\begin{eqnarray}
v_\Vk(x_{\IId}(x_\II)) &\hspace{-1.5mm}=\hspace{-1.5mm}& e^{-\beta\omega_\Vk\delta}\,v_\Vk(x_\II),
\label{vII}\\[2mm]
v^\sharp_{-\Vk}(x_{\IId}(x_\II)) &\hspace{-1.5mm}=\hspace{-1.5mm}&
e^{+\beta\omega_\Vk\delta}\,v_{-\Vk}^*(x_\II).
\label{vIIs}
\end{eqnarray}
We emphasize that the complex conjugation and the time shift do not commute.
Indeed, the mode $v^\sharp_{-\Vk}(x_{\IId}(x_\II))$ can be obtained from
$v_\Vk(x_{\IId}(x_\II))$ by complex conjugation {\it and} by the
replacement $\delta\rightarrow-\delta$. This rule actually defines the
$\sharp$-conjugation.

Similarly as in Eq.~(\ref{defphi}), one defines
\begin{eqnarray} \label{defphid}
\Phi(\xi)\ =\ \left\{
\begin{array}{lll}
\phi^\I(x_\I(\xi)), &\quad& \mbox{when $\xi\in R_\I$}, \\[2mm]
\phi^\IId(x_\IId(\xi)), && \mbox{when $\xi\in R_\IId$}.
\end{array}\right.
\end{eqnarray}
In $R_\I\cup R_\IId$ the Klein--Gordon-like inner product Eq.~(\ref{SP}) takes the form
\begin{equation}
(\Phi_1,\Phi_2)_{_\delta} \, = \,
\langle \phi^\I_1,\phi^\I_2\rangle \, + \, \langle \phi^\IId_1,\phi^\IId_2\rangle_{_\delta}\, ,
\end{equation}
where the local Minkowski inner product $\langle \ ,\ \rangle_{_\delta}$ in region
$R_\IId$ is given by
\begin{equation}
\langle \phi_1,\phi_2\rangle_{_\delta}\, = \, -i\int_{\R^3}\hspace{-2.2mm} d^3x_\IId\,
\phi_1(x_\IId)\stackrel{\leftrightarrow}{\partial}_t \phi_2^\sharp(x_\IId)\, .
\end{equation}
Equations (\ref{ACuv}), (\ref{ACvu}), (\ref{vII}) and (\ref{vIIs})
imply
\begin{eqnarray}
&&u_\Vk(x_\I(\xi_\delta))  \ = \
e^{-\beta\omega_\Vk(1/2+\delta)}\,v^\sharp_{-\Vk}(x_{\IId}(\xi_\delta))\, ,
\label{ACuvd} \\[2mm]
&&v_\Vk(x_\IId(\xi_\delta)) \ = \
e^{-\beta\omega_\Vk(1/2+\delta)}\,u^\sharp_{-\Vk}(x_\I(\xi_\delta))\, .
\label{ACvud}
\end{eqnarray}
With this, we can also write
\begin{eqnarray}
&&u^\sharp_\Vk(x_\I(\xi_\delta)) \ = \ e^{-\beta\omega_\Vk(1/2-\delta)}\,
v_{-\Vk}(x_{\IId}(\xi_\delta))\, ,
\label{ACuvds}\\[2mm]
&&v^\sharp_\Vk(x_\IId(\xi_\delta))  \ = \
e^{-\beta\omega_\Vk(1/2-\delta)}\,u_{-\Vk}(x_\I(\xi_\delta))\, .
\label{ACvuds}
\end{eqnarray}
The expressions on the left and right hand sides of these equations
are thus analytic continuations of each other. If we now define
\begin{eqnarray}
U_\Vk(\xid) &=&
\left\{\begin{array}{lcl}
u_\Vk(x_\I(\xid)), &\quad&
\mbox{when $\xid\in R_I$}, \\[2mm]
0, && \mbox{when $\xid\in R_\IId$},
\end{array}\right.
\\[3mm]
V_\Vk(\xid) &=&
\left\{\begin{array}{lcl}
0, && \mbox{when $\xid\in R_I$}, \\[2mm]
v_\Vk(x_{\IId}(\xid)), &\quad&
\mbox{when $\xid\in R_\IId$},
\end{array}\right.
\end{eqnarray}
the global modes in the extended Lorentzian section $R_\I\cup R_\IId$ are
be written as
\begin{eqnarray} \mlab{ex11}
\label{GM1}
&&G_{\Vk}(\xi_\delta) \ = \ (1\,-\,f_\Vk)^{-\frac{1}{2}} \left[ U_\Vk(\xi_\delta)
\,+\,f_\Vk^{\frac{1}{2}+\delta}\,V^\sharp_{-\Vk}(\xi_\delta) \right], 
\\[2mm]
&&{\widetilde G}_{\Vk}(\xi_\delta) \ = \
(1\,-\,f_\Vk)^{-\frac{1}{2}}\left[ V_\Vk(\xi_\delta) \,+\,
f_\Vk^{\frac{1}{2}+\delta}\,U^\sharp_{-\Vk}(\xi_\delta) \right],
\label{GM}
\end{eqnarray}
and
\begin{eqnarray} \mlab{ex11bis}
\label{GMs1}
&&G^\sharp_{\Vk}(\xi_\delta) \ = \
(1\,-\,f_\Vk)^{-\frac{1}{2}} \left[ U^\sharp_\Vk(\xi_\delta) \,+\,
f_\Vk^{\frac{1}{2}-\delta}\,V_{-\Vk}(\xi_\delta) \right],
\\[2mm]
&&{\widetilde G}^\sharp_{\Vk}(\xi_\delta) \ = \
(1\,-\,f_\Vk)^{-\frac{1}{2}} \left[ V^\sharp_\Vk(\xi_\delta)\,+\,f_\Vk^{\frac{1}{2}-\delta}\,U_{-\Vk}(\xi_\delta) \right],
\label{GMs}
\end{eqnarray}
where $f_\Vk\,=\,e^{-\beta\omega_\Vk}$ is the conventional Boltzmann factor. 

Equations~(\ref{ACuvd})-(\ref{ACvuds}) allow us to state that the global modes Eqs.~(\ref{GM1})-(\ref{GMs}) are analytic in $R_\I\cup
R_\IId$, in particular at the origin $\xi^+_\delta=\xi^-_\delta=0$.
Since they are analytic complex functions in the lower complex
planes of $\xi^+_\delta$ and $\xi^-_\delta$, their energy spectra have only
positive contributions. We might note that, for $\delta=0$, they reduce to the expressions
Eqs.~(\ref{globalmodes1})-(\ref{globalmodes}), as it should be expected.

Let us stress that the global modes $G^*_\Vk$ and
$\widetilde{G}^*_\Vk$ are {\em not} analytic in the extended
Lorentzian section, contrary to the non-Hermitian combinations
$G^\sharp_\Vk$ and $\widetilde{G}^\sharp_\Vk$.  Non-Hermitian
conjugation operations such as our ``sharp'' conjugation $\sharp$ are
actually common in TQFT, see for example Ref.~\cite{BESV98} for a
formally similar situation. In Ref.~\cite{OST}, the
necessity of the so-called Osterwalder--Schrader (reflection) positivity as
opposed to the Hermiticity property is shown in Euclidean field theories
even when the temperature vanishes.

The set $\{G_{\Vk},{\widetilde G}_{\Vk},G^\sharp_{\Vk}, {\widetilde
G}^\sharp_{\Vk}\}_{\Vk\in\R^3}$ is thus complete over $R_\I\cup
R_\IId$. It is furthermore an orthogonal set since
\begin{eqnarray}
&&(G_\Vk,G_\Vp)_\delta \ = \
(\widetilde{G}^\sharp_\Vk,\widetilde{G}^\sharp_\Vp)_\delta
\ = \ \hspace{-1.3mm} \delta^3(\Vk-\Vp)\,,  \\[2mm]
&&(G^\sharp_\Vk,G^\sharp_\Vp)_\delta \ = \
(\widetilde{G}_\Vk,\widetilde{G}_\Vp)_\delta
\ = \ -\delta^3(\Vk-\Vp)\, ,
\end{eqnarray}
with all the other inner products vanishing.

Following the procedure of Section~\ref{Lorentzian section}, we now expand the local fields in the Minkowski
modes over regions $R_\I$ and $R_\IId$ as
\begin{eqnarray} \mlab{expansionydelta1}
\phi^\I(x_\I) &\hspace{-1.5mm}=\hspace{-1.5mm}&
\int\hspace{-1mm} d^3k \left[ a^\I_\Vk\,u_\Vk(x_\I)\,+\,
a^{\I\dagger}_\Vk\,u^*_\Vk(x_\I)\right],
\\[1mm]\mlab{expansionydelta2}
\phi^\IId(x_\IId) &\hspace{-1.5mm}=\hspace{-1.5mm}&
\int\hspace{-1mm} d^3k \left[ a^\IId_\Vk\,v_\Vk(x_\IId)\,+\,
a^{_\IId\dagger}_\Vk\,v^*_\Vk(x_\IId)\right].
\end{eqnarray}
On the other hand, the expansion of the global field in the $G$ modes
over the region $R_\I\cup R_\IId$ reads as
\begin{equation}
\Phi(\xi_\delta) \,=\, \int\hspace{-1mm} d^3k\left[ c_\Vk\, G_\Vk(\xi_\delta)
\,+\, c^\sharp_\Vk\,G^\sharp_{\Vk}(\xi_\delta)
\,+\, {\tilde c}_\Vk\,{\tilde G}_{-\Vk}(\xi_\delta)\,
\,+\, {\tilde c}^\sharp_\Vk\,{\tilde G}^\sharp_{-\Vk}(\xi_\delta) \right].
\mlab{ex12}
\end{equation}
From these last expansions,  by using Eqs.~(\ref{GM1})-(\ref{GMs}), one finds the Bogoliubov transformations
\begin{eqnarray}
\label{BEc1}
c_\Vk & \hspace{-1mm}=\hspace{-1mm} & (1-f_\Vk)^{-\frac{1}{2}}
\left(a^\I_\Vk\,-\,f_\Vk^{\frac{1}{2}-\delta}\,a^{\IId\dag}_{-\Vk}\right),
\\[1.5mm]
{\tilde c}_\Vk & \hspace{-1mm}=\hspace{-1mm} & (1-f_\Vk)^{-\frac{1}{2}}
\left(a^\IId_{-\Vk}\,-\,f_\Vk^{\frac{1}{2}-\delta}\,a^{\I\dag}_\Vk\right),
\label{BEc}
\end{eqnarray}
and their $\sharp$-conjugate duals
\begin{eqnarray}
\label{BEcs1}
c^\sharp_\Vk &\hspace{-1mm} =\hspace{-1mm} & (1-f_\Vk)^{-\frac{1}{2}}
\left( a^{\I\dag}_\Vk \,-\, f_\Vk^{\frac{1}{2}+\delta}\,a^\IId_{-\Vk}\right),
\\[1.5mm]
{\tilde c}^\sharp_\Vk & \hspace{-1mm}=\hspace{-1mm} & (1-f_\Vk)^{-\frac{1}{2}}
\left( a^{\IId\dag}_{-\Vk}\,-\,f_\Vk^{\frac{1}{2}+\delta}\,a^\I_\Vk\right).
\label{BEcs}
\end{eqnarray}
Note that the Bogoliubov transformations Eqs.~(\ref{bogol})-(\ref{bogol2}) in the Lorentzian section are recovered
for $\delta\,=\,0$.

%%%%%%%%%%%%%%%%%%%%%%%%%%%%%%%%%%%%%%%%%%%%%%%%%%%%%%%%%%%
\section{Relationship between \EX spacetime and TQFTs}
%%%%%%%%%%%%%%%%%%%%%%%%%%%%%%%%%%%%%%%%%%%%%%%%%%%%%%%%%%%

We are now ready to formulate and to prove the connection
between QFTs in \EX spacetime and TQFTs. In particular, we will show
that in the aforesaid sections of \EX spacetime, QFT
naturally reproduces all the known formalisms of TQFTs inasmuch as the
correct thermal Green functions are recovered in respective sections.
Without loss of generality, we will carry out our discussion
in terms of a self-interacting real scalar field.

We start first by recalling the known result~\cite{GUI90} which states that in the Euclidean section of \EX
spacetime, QFT reproduces the imaginary time formalism. This can be seen both on the level
of generating functional and ensuing two-point Green functions. The latter turn out to be
nothing but thermal Green functions.
In the next step, we shall see that in the extended
Lorentzian section, QFT corresponds to the two known real-time TQFT formalisms,
namely the POM and TFD.
Moreover, we shall identify the parameter $\delta$ of the extended
Lorentzian section with the parameter $\sigma$ that naturally parametrizes both POM and TFD
formalisms. 

In order to show this, let us start from the general form of the Lagrange density in the full \EX spacetime for the real scalar theory with a Schwinger-type source term ${\rm J}$. This is given by
\begin{equation}
{\cal L}[\Phi,{\rm J}] \, = \, - \sqrt{g} \left(
\frac{1}{2}\,\partial_\mu\Phi\partial{\hspace{0.19mm}^\mu}\Phi \,+\,
\frac{m^2}{2}\,\Phi^2 \,+\, V(\Phi) \,-\, {\rm J} \Phi\right),
\end{equation}
where $V(\Phi)$ is an arbitrary local self-interaction which might be further restricted in its form, e.g. by requiring the renormalizability of the theory.

%%%%%%%%%%%%%%%%%%%%%%%%%%%%%%%%%%%%%%%%
\subsection{Matsubara formalism}
%%%%%%%%%%%%%%%%%%%%%%%%%%%%%%%%%%%%%%%%

In the Euclidean section of \EX spacetime, the generating functional of Green functions is
given by~\cite{GUI90}
\begin{equation}
Z_E[J] \, = \, N \int \hspace{-0.5mm}\mathcal{D}\Phi \exp\left\{-\hspace{-0.5mm}\int\hspace{-0.7mm} d\sigma d\xi dydz\
{\cal L}_{\sigma,\xi}[\Phi,{\rm J}]\right\},
\end{equation}
where
\begin{eqnarray}
\lefteqn{
{\cal L}_{\sigma,\xi}[\Phi,{\rm J}] \,=\,
\frac{1}{2}\,\left[\left(\partial_\sigma \Phi\right)^2
\,+\, \left(\partial_\xi \Phi\right)^2\right]
} \mlab{euclexz} \nonumber\\ [1mm]
 &&\hspace{3mm}
\hspace{10mm}+\, \frac{1}{\alpha^2\left(\sigma^2 \,+\,\xi^2\right)} \left\{
\frac{1}{2}\left(\nabla_{\!\perp}\Phi\right)^2
\,+\, \frac{m^2}{2}\,\Phi^2 \,+\, V(\Phi) \,-\, {\rm J}\Phi \right\},
\end{eqnarray}
is the corresponding pullback of the Lagrange density of the full \EX spacetime to the Euclidean section.

By performing the change of coordinates in Eqs.~(\ref{euclidTrans})-(\ref{euclidTrans2}), the generating functional takes the form
\begin{equation}\mlab{euclz}
Z_E[J] \, = \, N \int\hspace{-0.5mm} \mathcal{D}\Phi
\exp\left\{ \!-\!\int_0^\beta\!\!\!\!d\tau\hspace{-1.0mm}\int_{\R^3}\!\!\!\!
dx\hspace{0.1mm}dy\hspace{0.1mm}dz\,{\cal  L}_{\tau,x}[\phi,J] \right\},
\end{equation}
where the functional integration is taken over fields satisfying the Euclidean KMS boundary condition Eq.~(\ref{periodicfield}) and
\begin{equation}\mlab{euclzL}
{\cal L}_{\tau,x}[\phi,J]
\, = \, \frac{1}{2} \left[ \left( \partial_\tau\phi \right)^2
\, + \, \left( \nabla\phi \right)^2
\, + \, m^2\,\phi^2 \right] \,+\, V(\phi) \,-\, J\phi\, ,
\end{equation}
with $J(\tau,x,y,z)\,=\,{\rm J}(\sigma,\xi,y,z)$.
By differentiation of Eq.~(\ref{euclz}) with respect to the source $J$,
we obtain the Matsubara propagator, whose Fourier transform is
\begin{equation}
G_\beta({\bf k}, \omega_n) \, = \, \frac{1}{\omega_n^2 \,+\, {\bf k}^2 \,+\, m^2}\, .
\end{equation}
Here the (bosonic) Matsubara frequencies $\omega_n$ are given by
$\omega_n \,=\, 2\pi n/\beta$ ($n\in{\Bbb N}$).

%%%%%%%%%%%%%%%%%%%%%%%%%%%%%%%%%%%%%%%%%%%%%%%%%%%%%%
\subsection{Real time formalism -- POM}\label{Sec.4.2}
%%%%%%%%%%%%%%%%%%%%%%%%%%%%%%%%%%%%%%%%%%%%%%%%%%%%%%%

Let us now consider the extended Lorentzian section.
The generating functional of Green functions can be written as
\begin{equation}\mlab{lorentzexz}
Z[J] \, = \, {\cal N}\hspace{-0.5mm} \int \mathcal{D}\Phi \exp \left\{\!i\int\hspace{-0.7mm}
d\etad d\xid dydz\, {\cal L}_{\etad,\xid}[\Phi,{\rm J}] \right\},
\end{equation}
where
\begin{eqnarray}
\lefteqn{{\cal L}_{\etad,\xid}[\Phi,{\rm J}] \, = \,
\frac{1}{2} \left[ \left( \partial_{\etad}\!\Phi \right)^2
\,-\, \left( \partial_{\xid}\!\Phi \right)^2 \right]
} && \label{lorentzact} \\[1mm] \nonumber
&& \hspace{16mm}+ \, \frac{1}{\alpha^2\left|\xid^2\,-\,\etad^2\right|}
\left\{ - \frac{1}{2}\left(\nabla_{\!\perp}\Phi\right)^2
\,-\,\frac{m^2}{2}\Phi^2 \,-\, V(\Phi) \,+\, {\rm J}\Phi \right\}.
\end{eqnarray}
Since we are interested only in Green functions with spacetime
arguments belonging to $R_\I\cup R_\IId$, we can set the
source to zero in regions $R_\III$ and $R_\IV$, i.e.
${\rm J}(x)  \,=\,  0$ when $x\in R_\III\cup R_\IV$.
This amounts to reducing Eq.~(\ref{lorentzexz}) to
\begin{equation}\mlab{lorentzexz2}
Z[J] \, = \, {\cal N} \int\hspace{-0,5mm} \mathcal{D}\Phi \exp \left\{ i\int_{R_\I\cup R_\IId}
\hspace{-6mm} d\etad d\xid dydz\
{\cal L}_{\etad,\xid}[\Phi,{\rm J}] \right\}.
\end{equation}
By using the transformations in Eq.~(\ref{mex2a}), the fields in regions $R_\I$ and $R_\IId$
can now be expressed in terms of the local Minkowskian coordinates as
\begin{eqnarray}
&&\mbox{\hspace{-1.3cm}}Z[J] \ = \ {\cal N} \int\hspace{-0.5mm} \mathcal{D}\phi \ \! \exp \left\{i\int\hspace{-0.7mm} dt_\I dx_\I dydz \ \! {\cal L}_{t,x}[\phi,J](t_\I, {\bf x}_\I)\right\}
\nonumber \\[1mm]
&&\mbox{\hspace{0mm}} \times \ \! \exp \left\{i\int\hspace{-0.7mm}  dt_{\IId} dx_{\IId} dydz\ {\cal L}_{t,x}[\phi,J](t_{\IId}, {\bf x}_{\IId}) \right\},
\label{lorentzexz3}
\end{eqnarray}
where $\phi$ is the local field, the integration is taken over the Minkowski spacetime and
\begin{equation}\mlab{lorentzL}
{\cal L}_{t,x}[\phi,J]
%&\equiv& {\cal L}_0 [\phi,J]
%\label{L0} \\[2mm] \nonumber
\, = \, \frac{1}{2} \left[ \left( \partial_t\phi \right)^2
\,-\,  \left( \nabla\phi \right)^2
 \,-\,  m^2\,\phi^2 \right] - V(\phi) \,+\, J\phi\, .
\end{equation}
We now use Eqs.~(\ref{ACtxd1})-(\ref{ACtxd}) to further manipulate the generating
functional. It then follows that
\begin{eqnarray}
\label{lorentzexz4}
&&\mbox{\hspace{0mm}}Z[J] \, = \, {\cal N} \int \hspace{-0.5mm}\mathcal{D}\phi \ \! \exp
\left\{ i\int\hspace{-0.7mm} dt\hspace{0.3mm}dx\hspace{0.3mm}dy\hspace{0.3mm}dz
 \ \! {\cal L}_{t,x}[\phi,J]\,(t,{\bf x})\right\}
 \\[1mm]
\nonumber&&\mbox{\hspace{1.2cm}} \times \,  \exp \left\{- i\int\hspace{-0.7mm} dt\hspace{0.3mm}dx\hspace{0.3mm}dy\hspace{0.3mm}dz
 \ \! {\cal L}_{t,x}[\phi,J]\,\big(t \,+\, i\beta (1/2 \,+\, \delta), {\bf x}\big) \, \big] \right\}\hspace{-1mm},
\end{eqnarray}
where in the last step we have dropped the subscript $I$ and employed
the fact that the time direction (epitomized by time-like Killing vector) is mutually opposite in regions $R_\I$ and $R_\IId$.

Let us now consider the expression for the generating functional as given
in the POM formalism~\cite{LEB, Rivers}
\begin{equation}\mlab{POMgf}
Z_{\mbox{\tiny POM}}[J] \, = \, {\cal N}' \int\hspace{-0.5mm} \mathcal{D}\phi
\exp \left\{ i\int_C\hspace{-0.7mm} dt\hspace{0.3mm}dx\hspace{0.3mm}dy\hspace{0.3mm}dz\ {\cal L}_{t,x}[\phi,J]\right\}\, ,
\end{equation}
where the time path $C$ is the Niemi--Semenoff time path depicted in Fig.~\ref{fig1}. The 
path integration is over all fields satisfying periodicity condition $\phi(t_i, \textbf{x})\,=\,\phi(t_i\hspace{0.3mm}-\hspace{0.3mm}i\beta, \bf{x})$, $t_i$ being the initial time.
For most practical purposes (though not for all, see note in Discussion and Conclusions) one can disregard the contribution from 
the vertical parts of the path contour and assimilate it
in the normalization factor ${\cal N}'$~\cite{mabilat}.
In so doing, the generating functionals Eqs.~(\ref{lorentzexz4})-(\ref{POMgf}) can be identified provided that
\begin{equation}
\mlab{sigma}
\delta  \, = \, \sigma \,-\, 1/2\, .
\end{equation}
Therefore, we see that the time path used in the POM formalism is directly related
to the ``rotation angle'' between the two regions $R_\I$ and
$R_\IId$.
%We thus have reproduced the class of time paths relevant for TQFT in a
%geometric way: different paths correspond to different values of the
%angle between the two regions $R_\I$ and $R_\IId$ of \EX spacetime.
From the quadratic sector (i.e., free-field part) of the Lagrangian in Eq.~(\ref{POMgf}), we can read-off
the thermal-matrix propagator, which in the momentum space acquires the familiar Mills form~\cite{DAS,LEB,LW87,Mills}
\begin{eqnarray}
\label{matprop1}
&&\Delta_{11}(k) \, = \, \displaystyle
\frac{i}{k^2-m^2+i0_+}\, +\, 2\pi\,n(k_0)\,\delta(k^2-m^2)\, ,
\\[3mm]
&&\Delta_{22}(k) \, = \, \Delta_{11}^*(k)\, ,
\\[4mm]
\label{matprop2}
&&\Delta_{12}(k) \, = \, e^{\sigma\beta k_0}
\left[\,n(k_0)+\theta(-k_0)\,\right] 2\pi\,\delta(k^2-m^2)\, ,
\\[4mm]
&&\Delta_{21}(k)\, = \, e^{-\sigma\beta k_0}
\left[\,n(k_0)+\theta(k_0)\,\right] 2\pi\,\delta(k^2-m^2)\, ,
\mlab{matprop}
\end{eqnarray}
where $n(k_0)\,=\,(e^{\beta |k_0|} \,-\,1)^{-1}$. It is worth noting that the parameter $\sigma$
explicitly appears only in the off-diagonal components of the matrix
propagator.

%%%%%%%%%%%%%%%%%%%%%%%%%%%%%%%%%%%%%%%%%%%%%%%%%%%%%%%%%%%%%%%%%%%%%%%%
\subsection{Real time formalism -- TFD \label{Sec.4.3}}
%%%%%%%%%%%%%%%%%%%%%%%%%%%%%%%%%%%%%%%%%%%%%%%%%%%%%%%%%%%%%%%%%%%%%%%%

As already mentioned in Introduction, there is yet another formalism for real-time TQFT, namely
Thermo Field Dynamics~\cite{BJV,UM0,UM01,UM1,Nair}.
In this approach, a crucial r\^{o}le is played by the
Bogoliubov transformation relating the zero-temperature
annihilation and creation operators with the thermal ones.  In TFD
the field algebra is doubled and one then considers two commuting
field operators $\phi$ and ${\tilde \phi}$ given by
\begin{eqnarray}
&& \phi(x)\,=\,\int \hspace{-1mm}
\frac{d^{3}{k}}{(2\pi)^{\frac{3}{2}}
(2\omega_k)^{\frac{1}{2}}}\,
\left[a_{{\bf k}}e^{i\left(-\omega_k t \,+\, {\bf kx}\right)} \, + \,
a^{\dag}_{{\bf k}}e^{i\left(\omega_k t\,-\,{\bf kx}\right)}\right],
\mlab{4.21a} \\[2mm]
&&{\tilde \phi}(x) \, = \, \int \hspace{-1mm}
\frac{d^{3}{k}}{(2\pi)^{\frac{3}{2}}
(2\omega_k)^{\frac{1}{2}}}\,
\left[{\tilde a}_{{\bf k}}e^{i\left(\omega_k t\,-\,{\bf kx}\right)}\,+
\, {\tilde a}^{\dag}_{{\bf k}}
e^{i\left(-\omega_k t \,+\,{\bf kx}\right)}\right].
\label{4.21b}
\end{eqnarray}
The thermal Bogoliubov transformation in the
Takahashi--Umezawa representation~\cite{UM0,UM01,UM1} is given by
\begin{eqnarray}
\label{ex1}
a_\Vk(\theta) &\hspace{-1mm} =\hspace{-1mm} & a_\Vk\cosh\theta_\Vk \,- \,{\tilde
a}^{\dag}_\Vk\sinh\theta_\Vk\, , 
\\[2mm]
{\tilde a}_\Vk(\theta)
&\hspace{-1mm} =\hspace{-1mm} & {\tilde a}_\Vk\cosh\theta_\Vk \,-\, a^{\dag}_\Vk\sinh\theta_\Vk\, ,
\mlab{ex2}
\end{eqnarray}
where $\sinh^2\theta_{{\bf k}}\,=\,n(\omega_k) = (e^{\beta \omega_k}-1)^{-1}$. While the
operators $a_\Vk$ and ${\tilde a}_\Vk$ annihilate the (zero-temperature) vacuum $|0,{\tilde 0}\rangle \,=\, |0\rangle \otimes |{ 0}\rangle$,
the operators $a_\Vk(\theta)$ and ${\tilde a}_\Vk(\theta)$ 
annihilate the so-called ``thermal vacuum''
\begin{equation} 
|0(\beta)\rangle \, = \, \sum_{m}\rho^{\hspace{0.1mm}1/2}(\beta)|m,\tilde{m}\rangle\, ,
\label{thermvac.a}
\end{equation}
% Thermal averages
% are calculated as expectation values with respect to
% $|0(\beta)\rangle$.
%
where $|m,\tilde{m}\rangle \,\equiv\, |m\rangle\otimes |{m}\rangle$ with $|m\rangle$  being base vectors of the Fock space 
in the (regularized) occupation number representation, i.e. $|m\rangle \,\equiv\, |m_{\Vk_1}, m_{\Vk_2}, \ldots \rangle$. 
A doubled Fock space spanned by $|m,\tilde{m}\rangle$ is known as the Liouville space~\cite{BJV,H95}.
The operator $\rho(\beta)$ is the density matrix
defined by
\begin{equation}
\label{eqn:densmat}
\rho(\beta) \, = \, c \prod_{\bf{k}}f_{\bf{k}}^{\hspace{0.4mm}a^\dagger_{\bf{k}}a_{\bf{k}}} \otimes  1  \mbox{\hspace{-1mm}} {\rm{I}}  \, .
\end{equation}
Here $f_\Vk\,=\,e^{-\beta\omega_\Vk}$ is the conventional Boltzmann factor~\cite{footnoteII} and the coefficient $c$ is the normalization constant chosen so that 
$\langle 0(\beta)|  0(\beta)\rangle \,=\, 1$ holds. 

The form of the thermal Bogoliubov matrix, however, is not unique. Indeed, the
above transformations can be generalized to a non-Hermitian
superposition of the form~\cite{UM1,H95}
\begin{eqnarray}
\label{ex31}
\zeta_\Vk & \hspace{-1mm}= \hspace{-1mm}& (1\,-\,f_\Vk)^{-\frac{1}{2}}
\left(a_\Vk\,-\,f_\Vk^{1-\sigma}\,{\tilde a}^\dagger_\Vk \right), 
\\[2mm]
{\tilde\zeta}_\Vk & \hspace{-1mm}= \hspace{-1mm}& (1\,-\,f_\Vk)^{-\frac{1}{2}}
\left({\tilde a}_\Vk\,-\,f_\Vk^{1-\sigma}\,a^\dagger_\Vk \right).
\mlab{ex3}
\end{eqnarray}
The non-Hermitian property of
the last transformation implies that the canonical conjugates of
$\zeta$ and ${\tilde\zeta}$ are not $\zeta^\dagger$ and
${\tilde\zeta}^\dagger$, but are rather the combinations
\begin{eqnarray}
\label{ex41}
\zeta^\sharp_\Vk &\hspace{-1mm} =\hspace{-1mm} &(1\,-\,f_\Vk)^{-\frac{1}{2}}
\left( a^\dagger_\Vk \,-\,f_\Vk^\sigma\,{\tilde a}_\Vk \right), 
\\[2mm]
{\tilde\zeta}^\sharp_\Vk & \hspace{-1mm}=\hspace{-1mm} & (1\,-\,f_\Vk)^{-\frac{1}{2}}
\left( {\tilde a}^\dagger_\Vk \,-\,f_\Vk^\sigma\, a_\Vk \right),
\mlab{ex4}
\end{eqnarray}
which give the correct canonical commutators,
$[\zeta_\Vk,\zeta^\sharp_{\bf p}]\,=\,\delta^3({\bf k}-{\bf p})$, $[\zeta_\Vk,\zeta_{\bf p}]\,=\, 0$, $[\zeta^\sharp_\Vk,\zeta^\sharp_{\bf p}]
\,=\, 0$ and similarly for ${\tilde\zeta}_\Vk$ and ${\tilde\zeta}^\sharp_\Vk$.
Here the $\sharp$-conjugation is defined as the usual Hermitian
conjugation {\em together} with the replacement
$\sigma\rightarrow 1-\sigma$. The Hermitian representation Eqs.~(\ref{ex1})-(\ref{ex2}) is recovered when $\sigma \,=\, 1/2$.
Thermal averages are now expressed as~\cite{BJV,UM1,H95,Nair}
%
%\begin{equation}\mlab{4.31}
%\langle A \rangle =
%\frac{\,_{_L}\!\langle 0(\beta)| A | 0(\beta)\rangle_{_R}}
%{\,_{_L}\!\langle 0(\beta)|0(\beta)\rangle_{_R}},
%\end{equation}
%
\begin{equation}\mlab{4.31}
\langle A \rangle  \, = \,
\frac{(\hspace{-0.9mm}( \rho^L|\hspace{-0.5mm}| A |\hspace{-0.5mm}| \rho^R)\hspace{-0.9mm})}{(\hspace{-0.9mm}(\rho^L|\hspace{-0.5mm}|\rho^R)\hspace{-0.9mm})},
\end{equation}
where $A$ is a generic operator acting on the Liouville space and
\begin{equation}
\label{eqn:vacua}
|\hspace{-0.5mm}|\rho^R)\hspace{-1mm}) \, = \, 
\exp\left(\prod_{\bf{k}}f_{\bf{k}}^{\hspace{0.4mm}\sigma}\hspace{0.2mm}a^\dagger_{\bf{k}}\tilde{a}^\dagger_{\bf{k}}\right)\hspace{-1mm}|0,\tilde{0}\rangle, 
\quad\,\,\,
(\hspace{-1mm}(\rho^L|\hspace{-0.5mm}| \, = \, 
\langle 0,\tilde{0}|\exp\hspace{-1mm}\left(\prod_{\bf{k}}f_{\bf{k}}^{\hspace{0.4mm}(1-\sigma)}\hspace{0.2mm}a_{\bf{k}}\tilde{a}_{\bf{k}}\right)\hspace{-1mm}.
\end{equation}
In the special case when $A \,\equiv\, A\otimes 1  \mbox{\hspace{-1mm}} {\rm{I}}$, then $\langle A \rangle$ reduces to the standard thermal average of an observable $A$.
Again, for $\sigma\,=\,1/2$, the states $|\hspace{-0.5mm}|\rho^R)\hspace{-1mm})$ and $(\hspace{-1mm}(\rho^L|\hspace{-0.5mm}|$ become Hermitian conjugates. 
Furthermore, by employing Eqs.~(\ref{ex31})-(\ref{ex4}) and (\ref{eqn:vacua}), it can be seen that
\begin{equation}\mlab{4.35}
\left.\begin{array}{c}
\zeta\\{\tilde \zeta}
\end{array}\right\}\hspace{-0.5mm}
|\hspace{-0.5mm}|\rho^R)\hspace{-1mm})\,=\,0 \,=\,
(\hspace{-1mm}(\rho^L|\hspace{-0.5mm}|
\left\{\begin{array}{c}
\zeta^{\sharp} \\{\tilde \zeta}^{\sharp}
\end{array}\right..
\end{equation}
The thermal propagator for a scalar field in TFD is calculated as
\begin{equation}\mlab{4.36}
\Delta_{ab}(x,y) \, = \, 
\langle T\left[ \phi^a(x) \phi^{b\dagger}(y) \right] \rangle\, ,
\end{equation}
where $T$ is the time ordering symbol and the $a,b$ indices refer to the thermal doublet $\phi^1 \,= \,\phi$ and $\phi^2 \,=\, \tilde{\phi^{\dagger}}$. In the present case of a 
real scalar field we should use in the above definition $\phi^2 \,=\, \tilde{\phi}$.
Quite remarkably, the propagator Eq.~(\ref{4.36}) is equal to the one given by Eqs.~(\ref{matprop1})-(\ref{matprop}), as it can
be easily verified by employing the definitions given above.  

The connection of TFD with the geometric picture of \EX spacetime is
immediate by making the identification
\begin{equation}
\hspace{-1mm}\left(\begin{array}{c}
\phi \\ {\tilde \phi}
\end{array}\right)
\, \equiv \,
\left(\begin{array}{c}
\phi^\I \\ \phi^\IId
\end{array}\right).
\end{equation}

Let us now analyze some other salient features of \EX spacetime, which
are directly related the rotation Eqs.~(\ref{mex31})-(\ref{mex3}).  Along the lines
of Ref.~\cite{ZG98}, we consider the analytic extension of the
imaginary time thermal propagator to real times within the framework of \EX
spacetime. In Ref.~\cite{ZG98} it was shown that the geometric
structure of this spacetime plays a central role in obtaining the
matrix real-time propagator from the Matsubara one.
In order to see how this works, we consider the simple case of
a massless free scalar field in two-dimensions. In the Euclidean
section, the equation for the propagator has the form
\begin{equation}\mlab{of1}
\left(\frac{\partial^2}{\partial \sigma^2}
\,+\,\frac{\partial^2}{\partial \xi^2} \right)\hspace{-0.2mm}\Delta_E(\xi^{\mu}\,-\,\xi^{\mu'})
\, = \, - (g_E)^{-\frac{1}{2}}\delta(\xi^{\mu}\,-\,\xi^{\mu'})\, ,
\end{equation}
where $(\xi^{\mu}, \xi^{\mu'})$ denotes a couple of points, $g_E$ stands for the determinant of
the pullback metric in the Euclidean section and $\Delta_E$ is the
imaginary time thermal propagator. Let us now continue Eq.~(\ref{of1}) to the
extended Lorentzian section. This is achieved by first replacing
$\sigma$ by $-i\eta$ and then performing the rotation in Eqs.~(\ref{mex31})-(\ref{mex3}). If $\Delta$ is the real time propagator, we have
\begin{equation}\mlab{of2}
\left(-\frac{\partial^2}{\partial \etad^{2}} \,+\,
\frac{\partial^2}{\partial \xid^{2}} \right)\hspace{-0.2mm}\Delta(\xi^\mu\,-\,\xi^{\mu'})
\,=\, - (g_{L})^{-\frac{1}{2}}\delta(\xi^\mu\,-\,\xi^{\mu'})\, ,
\end{equation}
where $g_{L}$ is the absolute value of the determinant of the pullback metric in the
Lorentzian section. Due to the presence of different disconnected
regions in the Lorentzian section, the propagator exhibits a matrix
structure, since now the points $\xi^\mu$ and $\xi^{\mu'}$ can belong either to
region $R_{\I}$ or $R_{\IId}$ ($R_{\III}$ and $R_{\IV}$, as mentioned earlier, are excluded
from our consideration). 
In terms of the Minkowski coordinates, Eq.~(\ref{of2}) reads
\begin{equation}\mlab{of3}
\left(-\frac{\partial^2}{\partial t^{2}}
\, + \, \frac{\partial^2}{\partial x^{2}} \right) \hspace{-0.2mm}\Delta(\xi^{\mu} \,-\, \xi^{\mu'})
\, = \, - \delta_C(\xi^{\mu} \,-\, \xi^{\mu'}),
\end{equation}
where the $\delta_C$ is defined as derivative of a contour step function  
\begin{equation}
\theta_C(t-t')\,=\,\theta(s-s'),
\end{equation}
so that 
\begin{equation}
\delta_C(t-t')\, =\, {\left(\frac{dz}{ds}\right)}^{-1}\delta(s-s').
\end{equation}
Here $t\,=\,z(s)$, with $s\in \mathbb{R}$ monotonically increasing parameterization of the time path $C$.
This path coincides with the Niemi--Semenoff time path of Fig.~\ref{fig1}
when the identification Eq.~(\ref{sigma}) is made.  By using Eqs.~(\ref{ACtxd1})-(\ref{ACtxd}),
we obtain, for example, for the component $\Delta_{12}$, the equation
\begin{equation}
{ \left(-\frac{\partial^2}{\partial t^{2}}
\, + \, \frac{\partial^2}{\partial x^{2}}\right)\hspace{-0.2mm}\Delta(t\hspace{0.5mm}-\hspace{-0.5mm}t'\hspace{0.5mm}+\hspace{0.5mm}i\sigma\beta,x\hspace{0.5mm} -\hspace{0.5mm}x') }
\, = \, - \delta_C(t\hspace{0.5mm}-\hspace{0.5mm}t'\hspace{0.5mm}+\hspace{0.5mm}i\sigma\beta) \,\delta(x \hspace{0.5mm}-\hspace{0.5mm}x'),
\mlab{of5}
\end{equation}
which gives us the solution propagator $\Delta_{12}$ given in Eq.~(\ref{matprop2}).

We finally consider the {\em tilde conjugation} within the framework of the
\EX spacetime. The tilde conjugation rules are postulated in TFD in
order to connect the physical and the tilde operators.  Due to the
geometric structure of \EX spacetime, these rules are there seen as
coordinate transformations. This result, which was first discussed in
Ref.~\cite{ZG95}, is here enlarged to the extended Lorentzian
section of \EX spacetime.

Let us recall the tilde conjugation rules as originally defined in TFD \cite{UM0,UM01,UM1} (we
restrict for simplicity to bosonic operators):
\begin{eqnarray}
\begin{array}{rclcrcl}
\left(A B\right)\tilde{} &\hspace{-1.5mm}=\hspace{-1.5mm}& \tilde{A} \tilde{B}, &&\quad
\left(c_1 A \,+\, c_2 B \right)\tilde{} &\hspace{-1.5mm}=\hspace{-1.5mm}& c_1^*\tilde{A} \,+\, c_2^*\tilde{B}\, ,
\\
\left(\tilde{A}\right)\tilde{} &\hspace{-1.5mm}=\hspace{-1.5mm}& A, &&\quad
\hspace{0.2mm}\left(\tilde{A}\right)^\dagger &\hspace{-1.5mm}=\hspace{-1.5mm}& \left(A^\dagger\right)\tilde{},
\end{array}\mlab{of6}
\end{eqnarray}
where $A, B$ are two generic operators and $c_1, c_2$ are $c$-numbers.  In order to
reproduce this operation in the extended Lorentzian section, let us first introduce the following $M$ operation as defined in Ref.~\cite{ZG95}:
\begin{equation}\mlab{of7}
M\,\Phi(\eta, \xi)\,M^{-1} \, \equiv \, \Phi(-\eta, -\xi)\, .
\end{equation}
By expressing the field in terms of the Minkowskian coordinates, the $M$ operation can be cast into
\begin{equation}\mlab{of8}
M \,\phi(t, x)\, M^{-1} \, = \, \phi(t\,-\,i\beta/2, x)\, .
\end{equation}
Note that the $M$ operation is anti-linear, since it induces a time inversion
together with the shift $t\rightarrow t\,-\,i\beta/2$. This is clear when we consider its action on the
conjugate momentum $\pi(t,x)\,=\, \partial_t \phi^\dagger(t,x)$, then
\begin{equation}\mlab{of8b}
M\,\pi(t, x)\,M^{-1} \,= \, - \pi(t\,-\,i\beta/2, x)\, .
\end{equation}
Next we perform a rotation by an angle $\delta$ transforming the
$\eta,\xi$ coordinates according to Eqs.~(\ref{mex31})-(\ref{mex3}). The field then
becomes
\begin{equation}\mlab{of8c}
R_\delta\,\Phi(\eta,\xi)\,R_\delta^{-1} \, \equiv \, \Phi(\etad,\xid)\, .
\end{equation}
Finally, we introduce a $\delta$-conjugation operation, which is similar to a charge conjugation, by
\begin{equation}\mlab{of9}
C_\delta\,\phi(t, x)\,C_\delta^{-1} \, \equiv \, \phi^\sharp(t, x)\, .
\end{equation}
Here the change $\delta\rightarrow -\delta$ (or equivalently
$\sigma\rightarrow 1-\sigma $) has to be performed together with
usual charge conjugation.

The combination of these three operations results in the tilde
conjugation. By defining the combined transformation $G_\delta \,\equiv\,
C_\delta\,R_\delta\,M$, we have
\begin{equation}\mlab{o10}
G_\delta\,\phi(t, x)\,G_\delta^{-1} \, = \,  \phi^\sharp(t\,-\,i\sigma \beta, x)
\,= \, \phi^\dagger(t\,-\,i(1\,-\,\sigma)\beta,x)\, .
\end{equation}
In order to reproduce the tilde rules of Eq.~(\ref{of6}),  we can now omit for simplicity the space dependence of the field. Then we have
\begin{eqnarray}
&&\mbox{\hspace{-10mm}}G_\delta\, \phi_1(t)\,\phi_2(t')\,G_\delta^{-1}
\,= \, \phi_1^\sharp(t\,-\,i\sigma \beta)\,\phi_2^\sharp(t'\,-\,i\sigma\beta)\, ,
\\[2mm]
&&\mbox{\hspace{-10mm}}G_\delta \left[G_\delta\,\phi(t)\,G_\delta^{-1}\right] G_\delta^{-1}
\, = \, \phi(t)\, ,
\\[2mm] \mlab{of11}
&&\mbox{\hspace{-10mm}}G_\delta \left[ B_1\,\phi_1(t) \, +  \, B_2\,\phi_2(t') \right]
G_\delta^{-1} \, = \,
B_1^*\,\phi_1^\sharp(t\,-\,i\sigma \beta) \, + \,
B_2^*\,\phi_2^\sharp(t'\,-\,i\sigma\beta)\, ,
 \\[2mm]
&&\mbox{\hspace{-10mm}}G_\delta\left[ \phi^\dagger(t) \right] G_\delta^{-1}
 \, = \, \left[ G_\delta\,\phi(t)\,G_\delta^{-1}\right]^\dagger.
\end{eqnarray}
The $c$-numbers are conjugated since the $M$ operation is
anti-linear. The second of the above relations follows form the fact that
\begin{eqnarray}
\mbox{\hspace{-6mm}}G_\delta \left[G_\delta \,
\phi(t)\, G_\delta^{-1}\right] G_\delta^{-1}
&\hspace{-1.5mm}=\hspace{-1.5mm}& C_\delta\,R_\delta\,M\,\phi^\dagger(t\,-\,i(1\,-\,\sigma) \beta)\,
M^{-1}\,R_\delta^{-1}\, C_\delta^{-1} \nonumber
\\[2mm]
&\hspace{-1.5mm}=\hspace{-1.5mm}& C_\delta\,\phi^\dagger(t\,-\,i(1\,-\,\sigma)\beta \,-\, i\sigma\beta)
\, C_\delta^{-1} \nonumber
\\[2mm] \mlab{of12}
&\hspace{-1.5mm}=\hspace{-1.5mm}& \ \phi(t\,-\,i\beta) \, = \, \phi(t)\, .
\end{eqnarray}
On the last line we have used the periodicity boundary condition for fields in the Lorentzian section, cf. Section~2.
In this way the tilde rules of Eq.~(\ref{of6}) for generic $\sigma$ are directly reproduced.

%%%%%%%%%%%%%%%%%%%%%%%%%%%%%%%%%%%%%%%%%%%%%%%%%%
\section{\bf Discussion and Conclusions}
%%%%%%%%%%%%%%%%%%%%%%%%%%%%%%%%%%%%%%%%%%%%%%%%%%

In this paper we have discussed the thermal properties of \EX spacetime.
Our particular focus was on specific complex sections of this spacetime
which could be identified with the general geometric background for real-time TQFTs
at equilibrium. More specifically, we have shown that there is a one-to-one relationship
between the {\em vacuum} Green's functions in the respective sections of \EX spacetime
and generic mathematical representation (the so-called Mills representation) of {\em thermal} Green's functions in Minkowski spacetime.
Complex sections discussed here can be regarded as an extension of the Lorentzian section
of Gui~\cite{GUI90} by means of a rotation of region $R_\II$ with respect
to $R_\I$ in the complex \EX spacetime. In terms of the Minkowski coordinates, this rotation is shown
to be an  isometry: it is equivalent to a constant time shift, leaving the metric invariant.
The angle between the two regions turns out to be related to the
$\sigma$ parameter of the time path as used in the POM formalism. It
also reproduces Umezawa's characteristic parameter appearing in the Bogoliubov thermal matrix
of TFD, when the relation between modes belonging to different regions
is considered.
%The general form of the thermal matrix propagator in the so-called Mills representation
%containing the $\sigma$ parameter has been obtained by use of this
%particular geometric background.
All in all, we have shown that the full \EX spacetime is versatile enough to allow for
analytic extension from the imaginary-time (Matsubara) propagator to generic $2\times 2$ thermal matrix
propagator  of the real-time formalism -- feat impossible in fixed 4-dimensional spacetime, and for a
consistent prescription of the tilde conjugation rule in TFD.

In the course of our analysis, we have seen that the geometric framework of \EX spacetime allows us to understand the various existent 
formalisms of TQFT in a unified way. In particular, with regard to the real-time
methods, i.e. the POM formalism and TFD, one can draw the following  geometric picture.
In the Lorentzian section of \EX spacetime there are two different regions $R_\I$ and $R_\IId$ over
which the global field is defined.  For a global observer this field propagates in a zero-temperature heat bath.
However, when one restricts itself to one of the two regions (say $R_\I$),
temperature arises as a consequence of the loss of information (increase in entropy) about the other region.
In order to calculate the corresponding thermal propagator, one needs then to compare the fields
defined in different regions. This can be done essentially in two conceptually distinct ways.

i) By  analytically continuing the field $\phi^\IId(x_\IId)$ defined in
region $R_\IId$ to region $R_\I$. In this case the time argument gets shifted
by $i\beta(1/2\,+\,\delta)$, as described in Section III. One thus ends up
with {\em one} field and  {\em two} possible time arguments, which can be
either $t$ or $t\,-\,i\beta(1/2\,+\,\delta) $. The generating functional
defined in \EX spacetime by following this procedure turns out to be
the same of the one defined in the POM formalism. The matrix structure of the two-point
thermal Green's function is obtained by functionally differentiating the generating functional
with respect to Schwinger sources with four possible combinations of time arguments.

ii) One can attach the information about the region to the field
operator rather than putting it in the time argument. In this case the
identification $\phi^\I(x)\,\equiv\, \phi(x)$ and $\phi^\IId(x)\,\equiv\,
{\tilde \phi}(x)$ can be made and one obtains the formalism of TFD,
which consists of {\em two} commuting field operators and a {\em single}
time argument. The matrix structure of the two-point thermal Green's function
arises then from the four possible combinations of {\em physical} and {\em tilde}
fields in the (thermal) vacuum expectation value.

These two pictures lead to the same physics which is manifested by the same matrix form of
thermal Green functions in generic Mills representation.
After all, this could be expected since both pictures represent just a different
``viewpoints'' of an inertial local observer in the context of full \EX spacetime.
In this connection it is interesting to mention the r\^{o}le of regions $R_{\III}$ and $R_{\IV}$  of the extended
Lorentzian section. These regions were intentionally omitted from our discussion in the main text.
This omission was motivated, partially by technical simplifications, but mainly by the fact that
for most applications  the vertical parts of the time path in POM can be neglected.
In particular, for computation of correlation functions in TQFT, the regions $R_{\I}$ and $R_{\IId}$ of
the extended Lorentzian section fully suffice. It is, however, known that when vacuum-bubble diagrams are
important (e.g., when vacuum pressure, effective action or Casimir effect are considered) then the full
Niemi--Semenoff POM with vertical time paths included is obligatory~\cite{mabilat}. In such cases
the partition function $Z_{\mbox{\tiny POM}}$  cannot be factorized (as assumed in Section~\ref{Sec.4.2}), i.e.
$Z_{\mbox{\tiny POM}} = Z_{R_{\I}\cup R_{\II}\cup R_{\III}\cup R_{\IV}} \neq Z_{R{\I}\cup R_{\II}} Z_{R_{\III}\cup R_{\IV}}$,
or in other words, the regions $R_{\III}$ and $R_{\IV}$  must be correlated with regions $R_{\I}$ and $R_{\II}$. Similar conclusion holds
also for the extended Lorentzian section. Note that the aforesaid cross-horizon correlation has purely quantum-mechanical origin
(presence of vacuum-bubble diagrams is required) and hence it cannot be explained by classical means.
One can estimate the correlation between $R_{\III} \cup R_{\IV}$ and $R_{\I} \cup R_{\II}$, for instance, by checking how much the vertical time paths
contribute  in observable quantities (such as pressure). 
In this connection the Casimir effect at finite temperature  with oscillating plates (or other geometries) is a particularly pertinent system. 
Conceptually a similar issue was recently considered in the context of Rindler spacetime in order to
explain the origin of Unruh radiation in terms of vacuum entanglement among
all four different regions of that spacetime~\cite{Higuchi:2017gcd}.

%This formalism can be finally useful in studies about entanglement of field in non-inertial frames~\cite{} and

%**** Although we initiated the discussion using thermal equilibrium, the idea o thermofield states
%can be used with any density matrix, so we ****

Another interesting question to ask is to what extend the connection between \EX spacetime and TQFTs
can be generalized to out-of-thermal-equilibrium situations. This would be highly desirable endeavor as in
the last two decades  there has been a demand for a new set of tools and concepts from QFT
to treat the non-equilibrium dynamics of relativistic many-body systems and for understanding of further ensuing issues
like dissipation, entropy, fluctuations, noise and decoherence in these systems.
The catalyst has been the infusion of (mostly) experimental data from nuclear particle physics in the relativistic
heavy-ion collision experiments (at LHC and RHIC), early-universe cosmology in the wake of high-precision
observations (such as WMAP, Planck probe or BICEP3), cold atom (such as Bose--Einstein) condensation physics in highly
controllable environments, quantum mesoscopic processes and collective phenomena in condensed matter systems
(topological insulators, spintronics or out-of equilibrium phase transitions), etc.

While the \EX spacetime connection can certainly be applied in the Linear-Response-Theory
(i.e., near-to-equilibrium situations), as there one still employs the real-time (equilibrium) thermal
Green functions and concomitant Keldysh--Schwinger (or Niemi--Semenoff) POM~\cite{LEB}, the situation far-from-equilibrium is considerably less clear.
The major difficulty that hinders the applicability of the outlined geometrical picture  to generic non-equilibrium QFT systems
is the lack of any (asymptotically time-like) Killing vector field in the geometry of a dynamical time-dependent spacetime.
This leave us without a preferred time coordinate with which we could study the problem. One possible way to proceed is to employ
the {Kodama vector} as a substitute for the Killing vector,
that it is parallel to the timelike Killing vector in the static case (as well as at spatial infinity if one
assumes the evolving spacetime is asymptotically flat)~\cite{helou,wiser}. This ``preferred'' time coordinate is also known as the 
Kodama time~\cite{wiser}. Whether this route can lead to a new conceptual paradigm remains yet to be seen. Work in the direction is presently under active investigation.

%%%%%%%%%%%%%%%%%%%%%%%%%%%%%%%%%%%%%%%%%%%%%%%%%%%%%%%%
\section*{Acknowledgments}
%%%%%%%%%%%%%%%%%%%%%%%%%%%%%%%%%%%%%%%%%%%%%%%%%%%%%%%

P.J.  was  supported  by the Czech  Science  Foundation under the Grant No. 17-33812L.
This work was also in part supported by the U.S. Army RDECOM - Atlantic Grant No. W911NF-17-1-0108.


\begin{thebibliography}{00}
\bibitem{footnoteII} 
The product $\prod_{\Vk}$ in Eq.~(\ref{eqn:densmat})
is unambiguously defined only at finite volume, in which case $\Vk$ is a discrete variable. Strictly speaking, the infinite volume limit we consider here can be performed
only after this product has been evaluated.
\end{thebibliography}
\end{document}